\documentclass{jfm}

\usepackage{graphicx}
\usepackage{natbib}
\usepackage{color}

\ifCUPmtlplainloaded \else
  \checkfont{eurm10}
  \iffontfound
    \IfFileExists{upmath.sty}
      {\typeout{^^JFound AMS Euler Roman fonts on the system,
                   using the 'upmath' package.^^J}%
       \usepackage{upmath}}
      {\typeout{^^JFound AMS Euler Roman fonts on the system, but you
                   dont seem to have the}%
       \typeout{'upmath' package installed. JFM.cls can take advantage
                 of these fonts,^^Jif you use 'upmath' package.^^J}%
      }
  \else
  \fi
\fi

\ifCUPmtlplainloaded \else
  \checkfont{msam10}
  \iffontfound
    \IfFileExists{amssymb.sty}
      {\typeout{^^JFound AMS Symbol fonts on the system, using the
                'amssymb' package.^^J}%
       \usepackage{amssymb}%
         \let\leq=\leqslant
         \let\geq=\geqslant
      }{}
  \fi
\fi

\ifCUPmtlplainloaded \else
  \IfFileExists{amsbsy.sty}
    {\typeout{^^JFound the 'amsbsy' package on the system, using it.^^J}%
     \usepackage{amsbsy}}
    {\providecommand\boldsymbol[1]{\mbox{\boldmath $##1$}}}
\fi

\providecommand\bnabla{\boldsymbol{\nabla}}
\providecommand\bcdot{\boldsymbol{\cdot}}

\newcommand\Rey{\mbox{\textit{Re}}}  

\newsavebox{\astrutbox}
\sbox{\astrutbox}{\rule[-5pt]{0pt}{20pt}}

\title[Modes and instabilities in magnetized spherical Couette flow]{Modes and instabilities\\in magnetized spherical Couette flow}

\author[A. Figueroa, N. Schaeffer, H.-C. Nataf and D. Schmitt]%
{A.\ns F\ls I\ls G\ls U\ls E\ls R\ls O\ls A\thanks{now at: Facultad de Ciencias, Universidad Aut\'onoma del Estado de Morelos, 62209, Cuernavaca Morelos, M\'exico. (alfil@uaem.mx)},
N.\ns S\ls C\ls H\ls A\ls E\ls F\ls F\ls E\ls R,
H.\ls-\ls C.\ns N\ls A\ls T\ls A\ls F\thanks{Email address for correspondence: Henri-Claude.Nataf@ujf-grenoble.fr},\ns
\and D.\ns S\ls C\ls H\ls M\ls I\ls T\ls T}

\affiliation{ISTerre, Universit\'e de Grenoble 1, CNRS, F-38041 Grenoble, France}

\pubyear{2012}
\volume{652}
\pagerange{119--126}
\date{?; revised ?; accepted ?. - To be entered by editorial office}
\begin{document}

\maketitle
\begin{abstract}
Several teams have reported peculiar frequency spectra for flows in a spherical shell.
To address their origin, we perform numerical simulations of the spherical Couette flow in a dipolar magnetic field, in the configuration of the $DTS$ experiment.
The frequency spectra computed from time-series of the induced magnetic field display similar bumpy spectra, where each bump corresponds to a given azimuthal mode number $m$.
The bumps show up at moderate Reynolds number ($\simeq 2\,600$) if the time-series are long enough ($>300$ rotations of the inner sphere).
We present a new method that permits to retrieve the dominant frequencies for individual mode numbers $m$, and to extract the modal structure of the full non-linear flow.
The maps of the energy of the fluctuations and the spatio-temporal evolution of the velocity field suggest that fluctuations originate in the outer boundary layer.
The threshold of instability if found at $\Rey_c = 1 \, 860$.
The fluctuations result from two coupled instabilities: high latitude B\"odewadt-type boundary layer instability, and secondary non-axisymmetric instability of a centripetal jet forming at the equator of the outer sphere.
We explore the variation of the magnetic and kinetic energies with the input parameters, and show that a modified Elsasser number controls their evolution.
We can thus compare with experimental determinations of these energies and find a good agreement.
Because of the dipolar nature of the imposed magnetic field, the energy of magnetic fluctuations is much larger near the inner sphere, but their origin lies in velocity fluctuations that initiate in the outer boundary layer. 

\end{abstract}

\section{Introduction}\label{sec:introduction}

It is now well established that the magnetic field of most planets and stars is generated by the dynamo mechanism \citep{Larmor19, elsasser46a}.
Motions within an electrically conducting medium can amplify infinitesimally small magnetic field fluctuations up to a level where the Lorentz force that results is large enough to stop their amplification.
This is possible for large enough values of the magnetic Reynolds number $Rm = U L/\eta$ (where $U$ is a typical flow velocity, $L$ a typical length, and $\eta$ is the magnetic diffusivity of the medium).

Analytical \citep{busse75} and numerical \citep{glatzmaier95a} convective dynamo models, in which the flow is driven by the buoyancy force of thermal or compositional origin, have demonstrated the relevance of the dynamo mechanism for generating the Earth's magnetic field.
Other forcings, due to precession, tides or impacts are also invoked to explain the fields of some other planets \citep{LeBars11}.

In year 2000, two experiments demonstrated dynamo action in the lab \citep{gailitis01, stieglitz01}.
In both cases, the forcing was mechanical, with a dominant large-scale flow.
Efforts to produce dynamo action with a highly turbulent flow are still on the way \citep{Lathrop11, Kaplan11, Frick10}, while a rich variety of dynamo behaviours have been discovered in the $VKS$ experiment \citep{berhanu07, monchaux07} when ferromagnetic disks stir the fluid.

All these experiments use liquid sodium as a working fluid. The magnetic Prandtl number $Pm = \nu/\eta$ of liquid sodium is less than $10^{-5}$ ($\nu$ is the kinematic viscosity), so that experiments that achieve $Rm$ of order 50 (as required for dynamo action) have kinetic Reynolds number $\Rey = U L/\nu$ in excess of $10^6$.
This contrasts with numerical simulations, which require heavy computations with $1024^3$ grid points to reach $\Rey=10^4$.
Since Reynolds numbers of flows in planetary cores and stars are much larger, we have to rely on theory to bridge the gap.
Dynamo turbulence is a crucial issue because dissipation is very much dependent upon the scale and strength of turbulent fluctuations.
The question of instabilities and turbulence is also central in the study of accretion disks \citep{Balbus91}.
Laboratory experiments can bring some constraints since they exhibit intermediate Reynolds numbers.

In that respect, the observation in several experiments of very peculiar frequency spectra, characterized by a succession of peaks or bumps deserves some attention.
Such bumpy spectra have been obtained in both spherical and cylindrical geometries, when rotation or/and magnetic fields are present, two ingredients that also play a major role in natural systems.

\citet{Kelley07} were the first to observe a bumpy spectrum in a rotating spherical Couette experiment.
A small axial magnetic field was applied and the induced field was used as a marker of the flow.
The authors showed that the frequency and pattern of the modes correspond to a set of inertial modes.
Inertial modes are the oscillatory linear response of a fluid to a time-dependent perturbation where the Coriolis force is the restoring force.
Several hypotheses have been put forward to explain the excitation of inertial modes in these experiments: overcritical reflection off the inner Stewartson layer \citep{kelley10}, and turbulence from the tangent cylinder on the inner sphere \citep{matsui11}.
Most recently, \citet{rieutord12} presented data recorded in the 3m-diameter spherical Couette experiment of Dan Lathrop's group at the University of Maryland, and proposed a new interpretation.
They stress that there is a critical Rossby number below which modes of a given azimuthal mode number $m$ are no longer excited, and show that this happens when the frequency of the mode is equal to the fluid velocity in the Stewartson layer above the equator of the spinning inner sphere.
This interpretation in terms of a critical layer opens new perspectives that need to be investigated in more detail.

\begin{figure}
	\begin{center}
		\includegraphics[width=5cm]{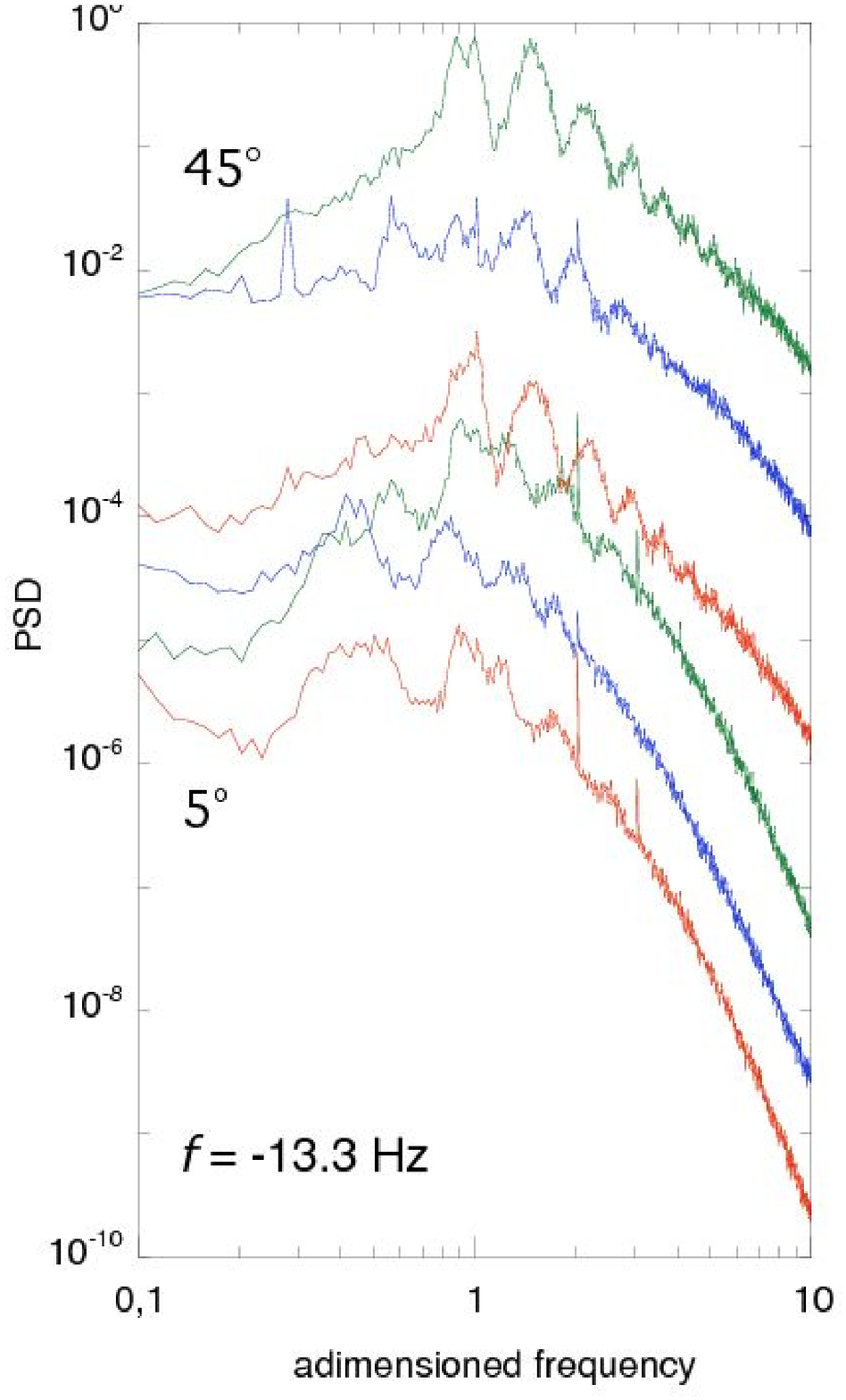}
		\includegraphics[width=6.5cm]{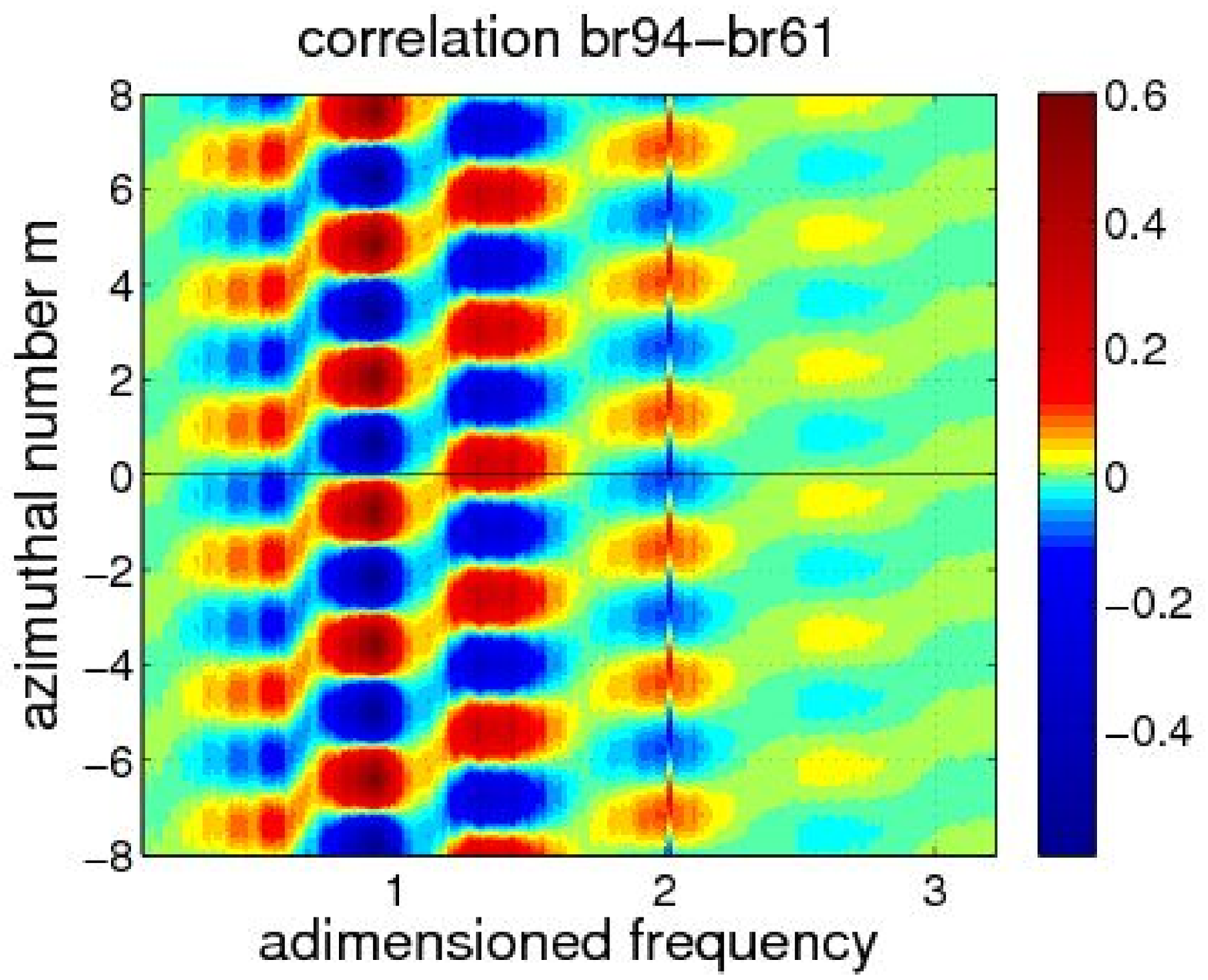}
	\end{center}
	\caption{Spectral bumps in the $DTS$ magnetized spherical Couette flow experiment.
a) Frequency spectra of the three components of the surface magnetic field ($b_r$, $b_\theta$, $b_\varphi$ in this order going up) measured at two different latitudes ($5^\circ$ and $45^\circ$ as indicated).
The time window used lasts $4 \, 000$ turns.
Frequencies ($x$-axis) are normalized by the rotation frequency of the inner sphere ($f = -13.3$ Hz).
The spectral energy density ($y$-axis) is normalized as in section \ref{subsec:energy_fluctuations} to facilitate the comparison with the corresponding simulation results.
The vertical scale applies to the lowest spectrum and successive spectra are shifted by one decade for clarity (two decades between different latitudes).
Note the succession of bumps that dominate the spectra.
b) covariance between two $b_r$ time-series recorded at points $128^\circ$ apart in longitude (same latitude $= -35^\circ$), in a frequency-azimuthal mode number ($m$) plot.
This plot shows that each spectral bumps in a) corresponds to a well-defined integer $m$, which increases with frequency (step-wise succession of positive covariance values starting at $m=0$).}
	\label{fig:DTS_modes}
\end{figure}

Bumpy frequency spectra were also reported by \citet{schmitt08} in the $DTS$ magnetized spherical Couette flow experiment \citep{Cardin02, Nataf06, Nataf08a, Brito11}.
An example is shown in figure \ref{fig:DTS_modes}a.
\citet{schmitt08} could show, by correlating signals measured at several longitudes, that each bump is characterized by a given azimuthal wavenumber $m$ (figure \ref{fig:DTS_modes}b).
\citet{schmitt12} further investigated the properties of the bumps and showed a good correspondence with linear magneto-inertial modes, in which both the Coriolis and the Lorentz forces play a leading role.

Finally, modes of azimuthal wavenumber $m=1$ were observed in two magnetized Couette flow experiments aimed at detecting the magneto-rotational instability (MRI): in spherical geometry in Maryland \citep{sisan04} and in cylindrical geometry in Princeton \citep{nornberg10}.
\cite{sisan04} discovered magnetic modes that appeared only when the imposed magnetic field was strong enough and interpreted their observations as evidence for the MRI, even though the most unstable mode is expected to be axisymmetric ($m=0$) in their geometry.
Rotating spherical Couette flow in an axial magnetic field was studied numerically by \citet{hollerbach09}, who suggested that instabilities of the meridional circulation in the equatorial region could account for some of the modes observed by \citet{sisan04}.
\citet{gissinger11} further investigated this situation and showed that the instabilities that affect the Stewartson layer around the inner sphere, modified by the imposed magnetic field, have properties similar to the MRI.
In contrast to the standard MRI, the instabilities evidenced by \citet{hollerbach09} and \citet{gissinger11} are inductionless.

In the cylindrical Taylor-Couette geometry with a strong imposed axial field, \citet{nornberg10} observed $m=1$ rotating modes.
They claimed that these modes could be identified to the fast and slow magneto-Coriolis waves expected to develop when both the Coriolis and Lorentz forces have a comparable strength.
Considering the fast magnetic diffusion in their experiment (Lundquist number of about 2), this interpretation was rather questionable, and indeed \citet{roach11} have recently reinterpreted these observations in terms of instabilities of an internal shear layer, in the spirit of the findings of \citet{gissinger11}.

Clearly, magnetized Couette flows display a rich palette of modes and instabilities, and it is important to identify the proper mechanisms in order to extrapolate to natural systems.
Hollerbach has investigated the instabilities of magnetized spherical Couette flow in a series of numerical simulations \citep{hollerbach01, hollerbach07, hollerbach09}.
However, bumpy spectra as observed by \citet{schmitt08} were never mentioned.
In this article, we perform numerical simulations in the geometry of the $DTS$ experiment, and focus on the origin of these bumpy spectra.
The observations of \citet{schmitt08} are illustrated by figure \ref{fig:DTS_modes}, but the reader should refer to their article for a more detailed presentation.
More specifically, we wish to answer the following major questions: how and where are the various modes excited? Are these spectra observed because of the large value of the Reynolds number?

We present the numerical model and the mean flow in section \ref{sec:numerical_model}. We perform spectral analyses in section \ref{sec:spectra}, and investigate fluctuations and instabilities in section \ref{sec:fluctuations}.
A discussion concludes the article.

\section{Numerical model and mean flow}\label{sec:numerical_model}

The $DTS$ experiment that we wish to model is a spherical Couette flow experiment with an imposed dipolar magnetic field.
Liquid sodium is used as a working fluid.
It is contained between an inner sphere and a concentric outer shell, from radius $r=r_i$ to $r=r_o$ ($r_i=74$ mm, $r_o=210$ mm).
The inner sphere consists of a $15$ mm-thick copper shell, which encloses a permanent magnet that produces the imposed magnetic field, whose intensity reaches 175 mT at the equator of the inner sphere.
The stainless steel outer shell is $5$ mm thick.
The inner sphere can rotate around the vertical axis (which is the axis of the dipole) at rotation rates $f=2 \pi \Omega$ up to $30$ Hz.
Although the outer shell can also rotate independently around the vertical axis in $DTS$, we only consider here the case when the outer sphere is at rest.

\begin{figure}
	\begin{center}
		\includegraphics[width=5.5cm]{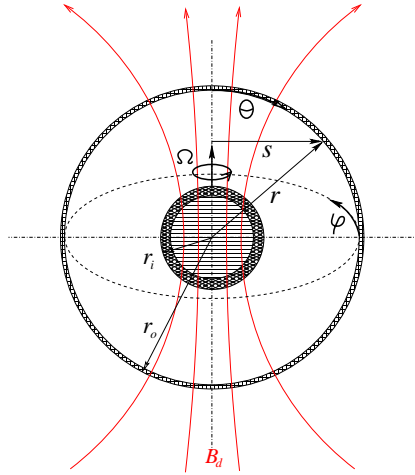}
	\end{center}
	\caption{Sketch of the $DTS$ experiment as modeled in this article. The inner sphere of radius $r_i$ rotates around the vertical axis at angular velocity $\Omega$. It consists of a copper shell enclosing a permanent magnet, which produces the imposed dipolar magnetic field $\boldsymbol{B}_d$. Liquid sodium of electric conductivity $\sigma$ fills the gap between the inner sphere and the stainless steel outer shell of inner radius $r_o$. The spherical coordinate system we use is drawn.}
	\label{fig:sketch}
\end{figure}

All these elements are taken into account in the numerical model, which is sketched in figure \ref{fig:sketch}.
In particular, we reproduce the ratio in electric conductivity of the three materials (copper, sodium, stainless steel).
In the experiment, the inner sphere is held by $25$ mm-diameter stainless steel shafts, which are not included in the numerical model.

\subsection{Equations}\label{subsec:equations}
We solve the Navier-Stokes and magnetic induction equations that govern the evolution of the velocity and magnetic fields of an incompressible fluid in a spherical shell:

\begin{equation}
	\frac{\partial \boldsymbol{ u}}{\partial t} + (\boldsymbol{ u} \bcdot \bnabla) \boldsymbol{ u}
=- \frac{1}{\rho} \bnabla p + \nu \nabla^2 \boldsymbol{ u} + \frac{1}{\mu_{0}\rho} (\bnabla \times \boldsymbol{ B}) \times \boldsymbol{ B},
	\label{eq:momentum_conservation}
\end{equation}

\begin{equation}
	\frac{\partial \boldsymbol{ B}}{\partial t} = \bnabla \times (\boldsymbol{ u} \times \boldsymbol{ B}) - \bnabla \times \left( \eta(r) \bnabla \times \boldsymbol{ B} \right),
	\label{eq:induction}
\end{equation}

\begin{equation}
	\bnabla\bcdot\boldsymbol{u} = 0,\quad \bnabla\bcdot\boldsymbol{ B} = 0,
	\label{eq:divergence_free}
\end{equation}
where $\boldsymbol{u}$ and $p$ stand for the velocity and pressure fields respectively.
Time is denoted by $t$, while $\rho$ and $\nu$ are the density and kinematic viscosity of the fluid.
The magnetic diffusivity $\eta(r)$ is given by $\eta(r) = (\mu_0 \sigma(r))^{-1}$ where $\sigma(r)$ is the electric conductivity of the medium (fluid or solid shells) and $\mu_0$ the magnetic permeability of vacuum. In the fluid, the conductivity $\sigma(r) = \sigma_{Na}$ is constant.
The last term of the first equation is the Lorentz force.
$\boldsymbol{B}$ is the magnetic field.
It contains the imposed dipolar magnetic field $\boldsymbol{B}_d$ given by:

\begin{equation}
	\boldsymbol{B}_{d} = B_{0} \frac{r_o^{3}}{r^{3}} \left( 2 \cos \theta \hspace*{0.05cm} \boldsymbol{e_r} + \sin \theta \hspace*{0.05cm} \boldsymbol{e_\theta} \right),
	\label{eq:dipole}
\end{equation}
\noindent where $\theta$ is the colatitude, $\boldsymbol{e_r}$ and $\boldsymbol{e_\theta}$ are the unitary vectors in the radial and orthoradial directions.
$B_0$ is the intensity of the field at the equator on the outer surface of the fluid ($r=r_o$).

\subsection{Boundary conditions}\label{subsec:boundary_conditions}

We use no-slip boundary conditions for the velocity field on the inner and outer surfaces:
\begin{equation}
	\boldsymbol{u} = \Omega r \sin\theta \boldsymbol{e_\varphi} \quad \mbox{for\ } \quad r \leq r_i,
 \quad \quad \boldsymbol{u} = \boldsymbol{0} \quad \mbox{for\ } \quad r \geq r_o.
	\label{eq:no-slip}
\end{equation}
We model the copper shell that holds the magnet in $DTS$ as a conductive shell with electric conductivity $\sigma_{Cu} = 4.2 \sigma_{Na}$.
The outer stainless steel shell is modelled as a shell of conductivity $\sigma_{SS} = \sigma_{Na}/9$.
These values reproduce the experimental conductivity contrasts.
The conductivity jumps are implemented by taking a continuous radial conductivity profile with sharp localized variations at both interfaces (3 to 5 densified grid points).
The internal magnet and the medium beyond the outer stainless steel shell are modelled as electric insulators.
The magnetic field thus matches potential fields at the inner and outer surfaces.

\subsection{Numerical scheme}\label{subsec:numerical_scheme}

Our three-dimensional spherical code (XSHELLS) uses second-order finite differences in radius and pseudo-spectral spherical harmonic expansion,
for which it relies on the very efficient spherical harmonic transform of the \texttt{SHTns} library \citep{Schaeffer12}.
It performs the time-stepping of the momentum equation in the fluid spherical shell, and the time-stepping of the induction equation both in the conducting walls and in the fluid.
It uses a semi-implicit Crank-Nicholson scheme for the diffusive terms, while the non-linear terms are handled by an Adams-Bashforth scheme (second order in time).
The simulations that we present typically have 600 radial grid points (with a significant concentration near the interfaces) while the spherical harmonic expansion is carried up to degree 120 and order 40.

\subsection{Dimensionless parameters}\label{subsec:dimensionless_parameters}

We define in table \ref{tab:numbers} the dimensionless numbers that govern the solutions in our problem.
We pick the outer radius $r_o$ as a length scale, and $B_0$, the intensity of the magnetic field at the equator of the outer surface, as a magnetic field scale.
Note that, due to the dipolar nature of the imposed magnetic field, its intensity is 23 times larger at the equator of the inner sphere.
The angular velocity of the inner sphere yields the inverse of the time scale.
We choose $U=\Omega r_i$, the tangential velocity at the equator of the inner sphere, as typical velocity.

\begin{table}
 	\begin{center}
		\def~{\hphantom{0}}
		\begin{tabular}{cccccc}
			symbol & expression & simulations & simulations & $DTS$ & Earth's\\
			& & fixed $Ha$ & fixed $Re$ & $f= [3-30]$ Hz & core\\[3pt]
			$Pm$ & $\nu/\eta$ & {$\mathbf{10^{-3}}$} & {$10^{-3}$} & {$7 \times 10^{-6}$} & {$10^{-6}$} \\
			$\Rey$ & $\Omega r_i r_o / \nu$ & $[\textbf{2\,611}-10\,100]$ & $2\,611$ &  {$[0.45 - 4.5] \times 10^6$} & $10^9$ \\
			$\Lambda$ &  $\sigma B_0^2/\rho \Omega$ & {$[\textbf{3.4} - 0.9] \times 10^{-2}$} & {$[0.21 - 13.1] \times 10^{-2}$} & $[3 - 0.3] \times 10^{-2}$ & 10 \\
			$Ha$ &  $r_o B_0/\sqrt{\rho \mu_0 \nu \eta}$ & $\textbf{16}$ & $[4 - 31]$ & $200$ & $10^8$
		\end{tabular}
		\caption{\label{tab:numbers}Typical values of the dimensionless numbers in the numerical simulations, in the $DTS$ experiment (computed for $f=\Omega/2 \pi = 3 - 30$ Hz) and in the Earth's core. The dimensionless numbers of our reference simulation are in bold.}
	\end{center}
\end{table}

The solutions are governed by three independent dimensionless numbers but several combinations are possible and we try to pick the most relevant ones.
The magnetic Prandtl number $Pm$ compares the diffusion of momentum to that of the magnetic field.
It is small in both the simulations and the experiment.
The Reynolds number $\Rey$ is of course essential, as it determines the level of fluctuations.
It is not feasible to run numerical simulations with Reynolds numbers as large as in the $DTS$ experiment.

However, one of the main findings of \citet{Brito11} is that, because of the imposed dipolar magnetic field, the time-averaged flow is mainly governed by the balance between the Lorentz and the Coriolis forces, where the latter is due to the global rotation of the fluid, which is very efficiently entrained by the inner sphere, even when the outer sphere is at rest.
That balance is measured by the Elsasser number $\Lambda$.
\citet{Brito11} showed that one can recover the proper balance at achievable values of $\Rey$ by reducing the influence of the magnetic field, keeping the effective Elsasser number $\Lambda$ as in the experiment.

Nevertheless, \citet{Cardin02} introduced another number $\lambda$ (named the Lehnert number by \citet{Jault08}), which provides a better measure of this balance for fast time-dependent phenomena.
The Lehnert number $\lambda$ compares the periods of Alfv\'en waves to that of inertial waves.
It is given by:
\begin{equation}
\lambda =  \frac{B_0}{\Omega r_i \sqrt{\rho \mu_0}}.
\end{equation}

In the Earth's core, this number is of order $10^{-4}$ and inertial waves dominate.
They force the flow to be quasi-geostrophic on short time-scales \citep{Jault08}.
However, magnetic diffusion severely limits the propagation of Alfv\'en waves in the $DTS$ experiment.
This is measured by the Lundquist number, which is the ratio of the magnetic diffusion time to the typical transit time of an Alfv\'en wave across the sphere, here given by:

\begin{equation}
Lu = \frac{r_o B_0}{\eta \sqrt{\rho \mu_0}},
\end{equation}
which is taken as $Lu = 0.5$ in the numerical simulations, in agreement with the experimental value.

We therefore follow the same strategy as \citet{Brito11}, and try to keep the Elsasser number of the numerical simulations similar to its experimental value.
Our reference case thus has: $Pm = 10^{-3}$, $\Rey = 2\,611$ and $\Lambda = 3.4 \times 10^{-2}$.
The Hartmann number is $Ha=16$, quite smaller that its experimental counterpart ($Ha=200$).
It follows that $\lambda=6.8 \times 10^{-2}$ and $Rm=\Rey Pm = 2.6$ for the reference case.
Most results shown in this article relate to our reference case, but we also present some results computed for other Reynolds and Hartmann numbers, as indicated in table \ref{tab:numbers}.

\subsection{Mean flow}\label{subsec:mean_flow}

\begin{figure*}
	\begin{center}
		\includegraphics[width=6.1cm]{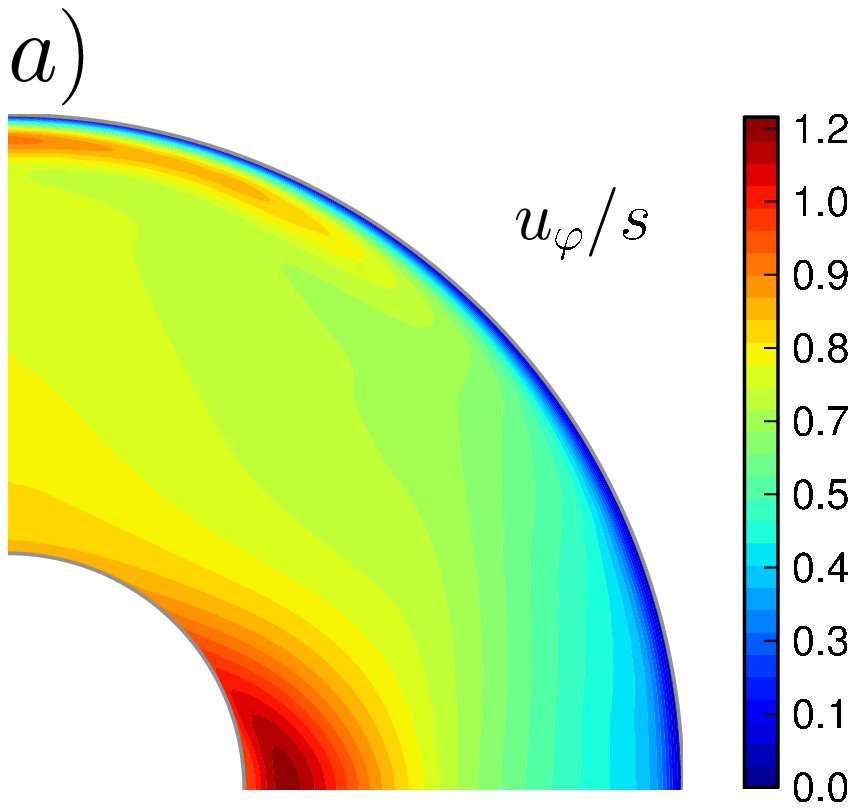}
\hspace*{1cm}
		\includegraphics[width=6.1cm]{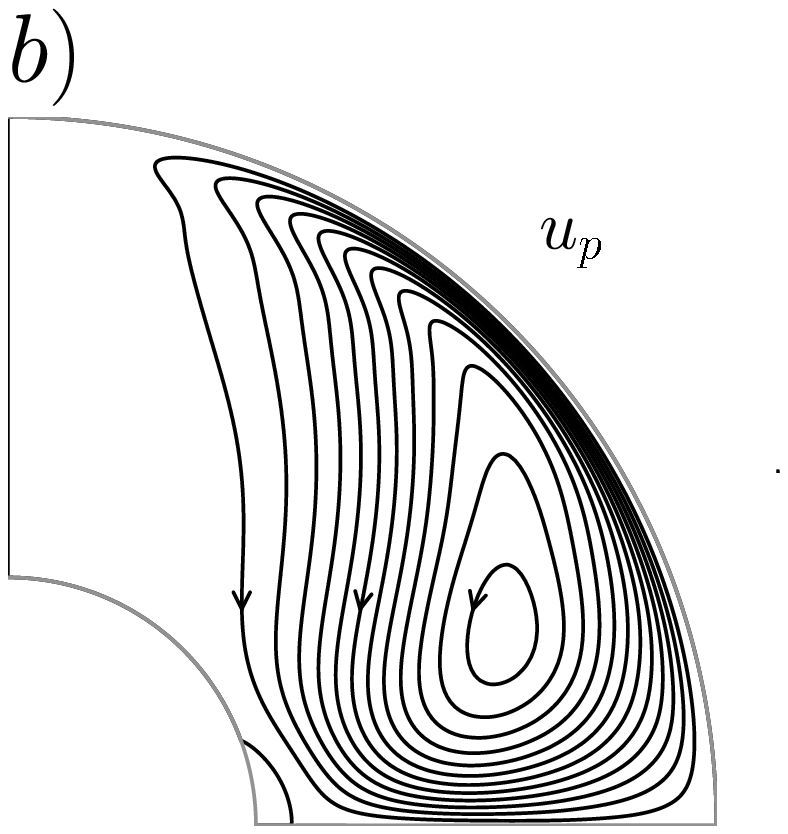}
	\end{center}
	\caption{Time-averaged meridional slice of the velocity field. a) Angular velocity isovalues. Note the zone of super-rotation near the inner sphere. b) Meridional streamlines. The maximum meridional velocity is 0.4. $Pm=10^{-3}$, $\Rey=2\,611$, $\Lambda=3.4 \times 10^{-2}$.}
	\label{fig:u_field_uw_upol_merid_Re_2600}
\end{figure*}

The time-averaged properties of the magnetized spherical Couette flow have been investigated in detail by \citet{Brito11}, and we simply recall here a few key observations.
We plot in figure \ref{fig:u_field_uw_upol_merid_Re_2600} the time-averaged velocity field in a meridional plane, for our reference simulation ($Pm=10^{-3}$, $\Rey=2\,611$ and $\Lambda=3.4 \times 10^{-2}$).
Two distinct regions show up in the map of mean angular velocity (figure \ref{fig:u_field_uw_upol_merid_Re_2600}a): an outer almost geostrophic region, where the angular velocity predominately varies with the cylindrical radius $s$; an inner region that tends to obey Ferraro law \citep{ferraro37} around the equator: the angular velocity is nearly constant along field lines of the imposed dipolar magnetic field.
Note the presence of a thin boundary layer at the outer surface.
The poloidal streamlines (figure \ref{fig:u_field_uw_upol_merid_Re_2600}b) display a circulation from the equator towards the poles beneath the outer surface, where the polewards velocity reaches $0.4 \, \Omega r_i $.

To check our numerical set-up, we compare the time-averaged velocity field of our simulation with that obtained by \citet{Brito11} using an independent axisymmetric equatorially-symmetric code.
The parameters and boundary conditions are identical, except that the magnetic boundary condition at $r=r_o$ is treated in the thin-shell approximation in \citet{Brito11}.

\begin{figure*}
	\begin{center}
		\includegraphics[width=6.5cm]{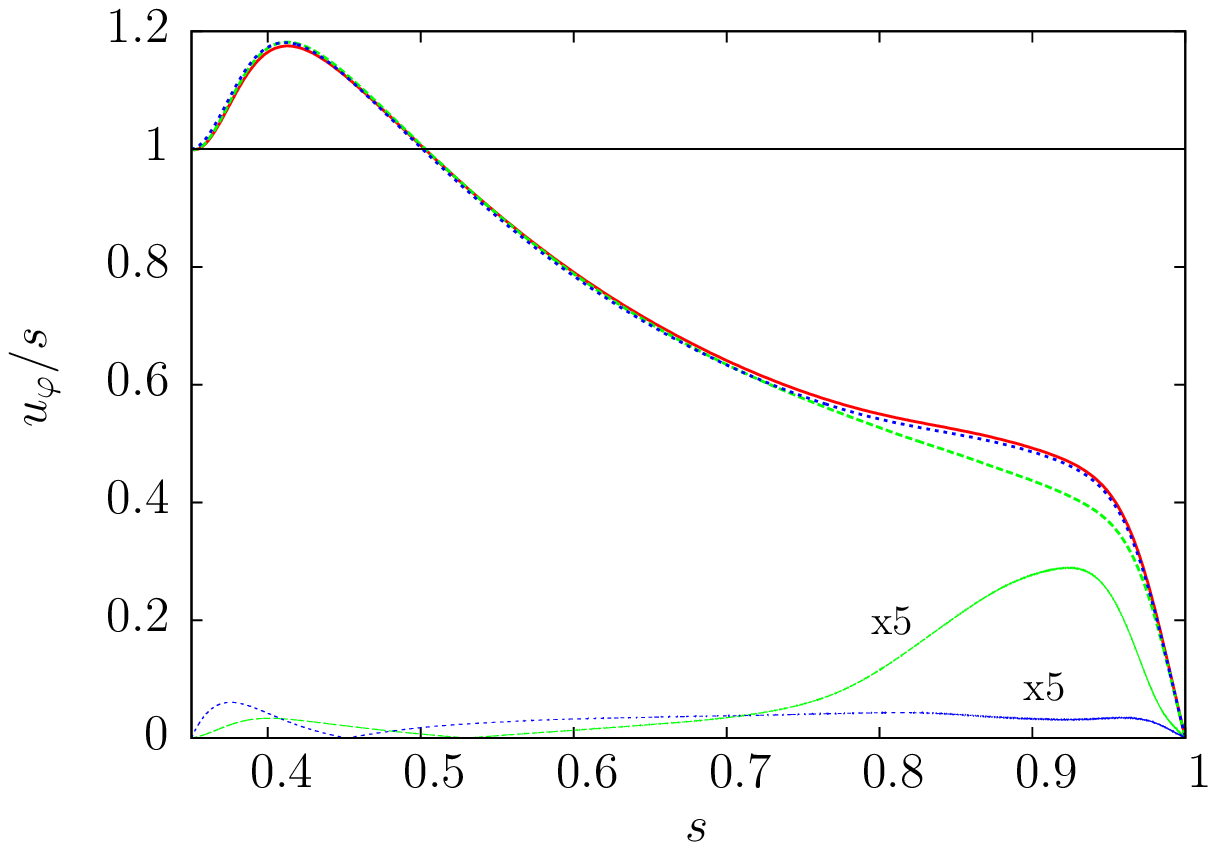}
	\end{center}
	\caption{Radial profiles of time-averaged angular velocity in the equatorial plane for different simulations with the same parameters ($Pm=10^{-3}$, $\Rey=2\,611$ and $\Lambda=3.4 \times 10^{-2}$). Dotted line: axisymmetric equatorially-symmetric solution of \citet{Brito11}; solid line: axisymmetric solution computed with our XSHELLS code; dashed line: 3D solution from XSHELLS. The curves at the bottom give the unsigned difference between the 3D solution and the axisymmetric one, and between the two axisymmetric solutions, scaled up by a factor 5.}
	\label{fig:mean_omega_profiles}
\end{figure*}

Figure \ref{fig:mean_omega_profiles} compares the radial profile of the angular velocity in the equatorial plane computed by \citet{Brito11} to our axisymmetric solution, averaged over $50$ rotation times, and to our 3D spherical solution, averaged over $100$ rotation times.
The two axisymmetric solutions agree very well, while the 3D solution exhibits a slightly lower angular velocity near the outer surface.
Note that the angular velocity of the fluid reaches values as high as $20\%$ larger than that of the inner sphere.
This phenomenon of {\it super-rotation} was first predicted by \citet{Dormy98} in the same geometry (also see \citet{Starchenko97}), but in their linear study, the zone of super-rotation was enclosed in the magnetic field line touching the equator of the outer sphere.
There, the induced electric currents have to cross the magnetic field lines in order to loop back to the inner sphere.
This produces a Lorentz force, which accelerates the fluid.
\citet{hollerbach07} showed that non-linear terms shift the zone of super-rotation from the outer sphere to close to the inner sphere, as observed here.
The excess of $20\%$ is in good agreement with the super-rotation measured in the $DTS$ experiment for $f = 3$ Hz ($\Rey = 4.5 \times 10^5$) \citep{Brito11}.

\section{Spectra and modes}\label{sec:spectra}

\subsection{Frequency spectra}\label{subsec:frequency_spectra}

\begin{figure*}
	\begin{center}
		\includegraphics[width=5cm]{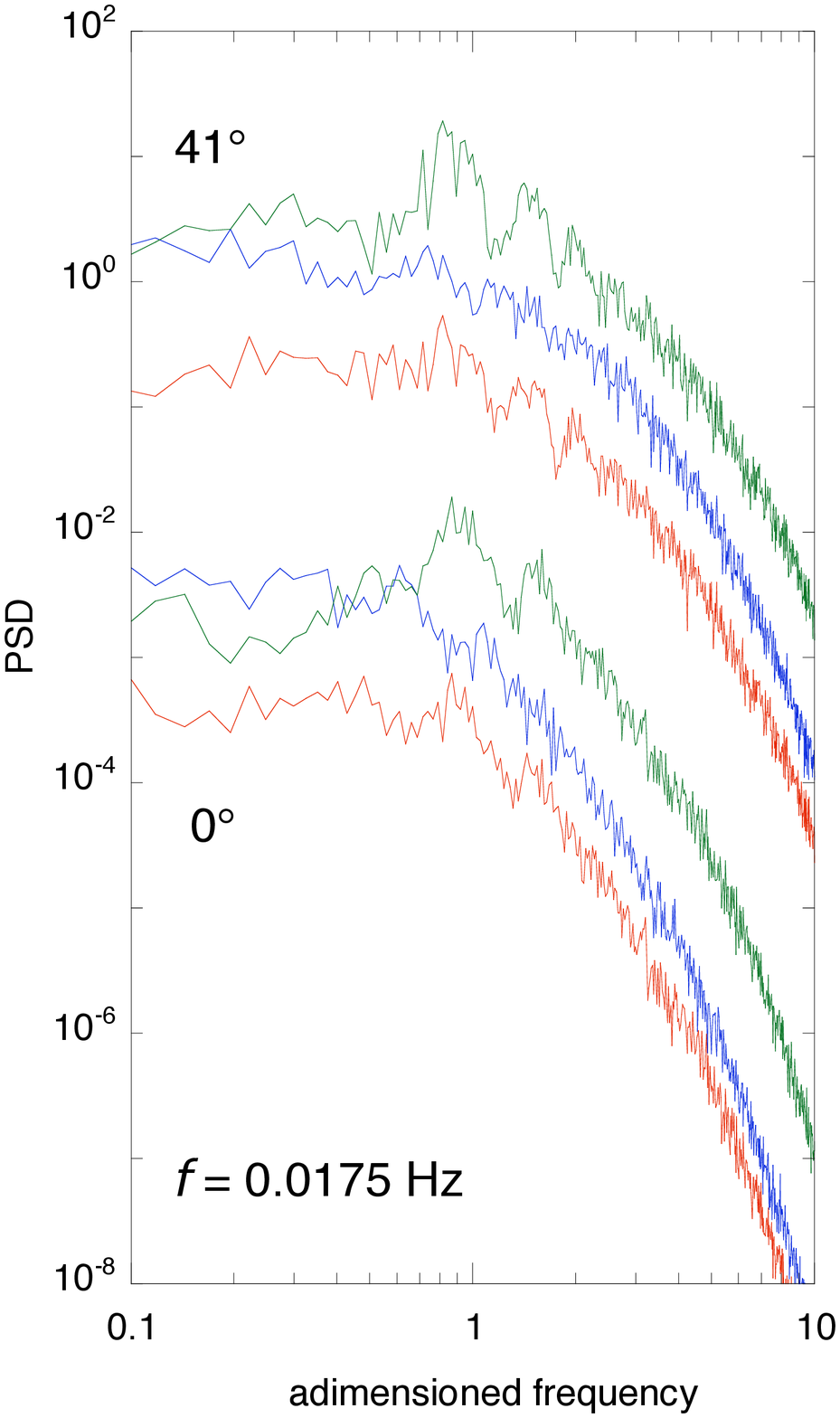}
		\includegraphics[width=6cm]{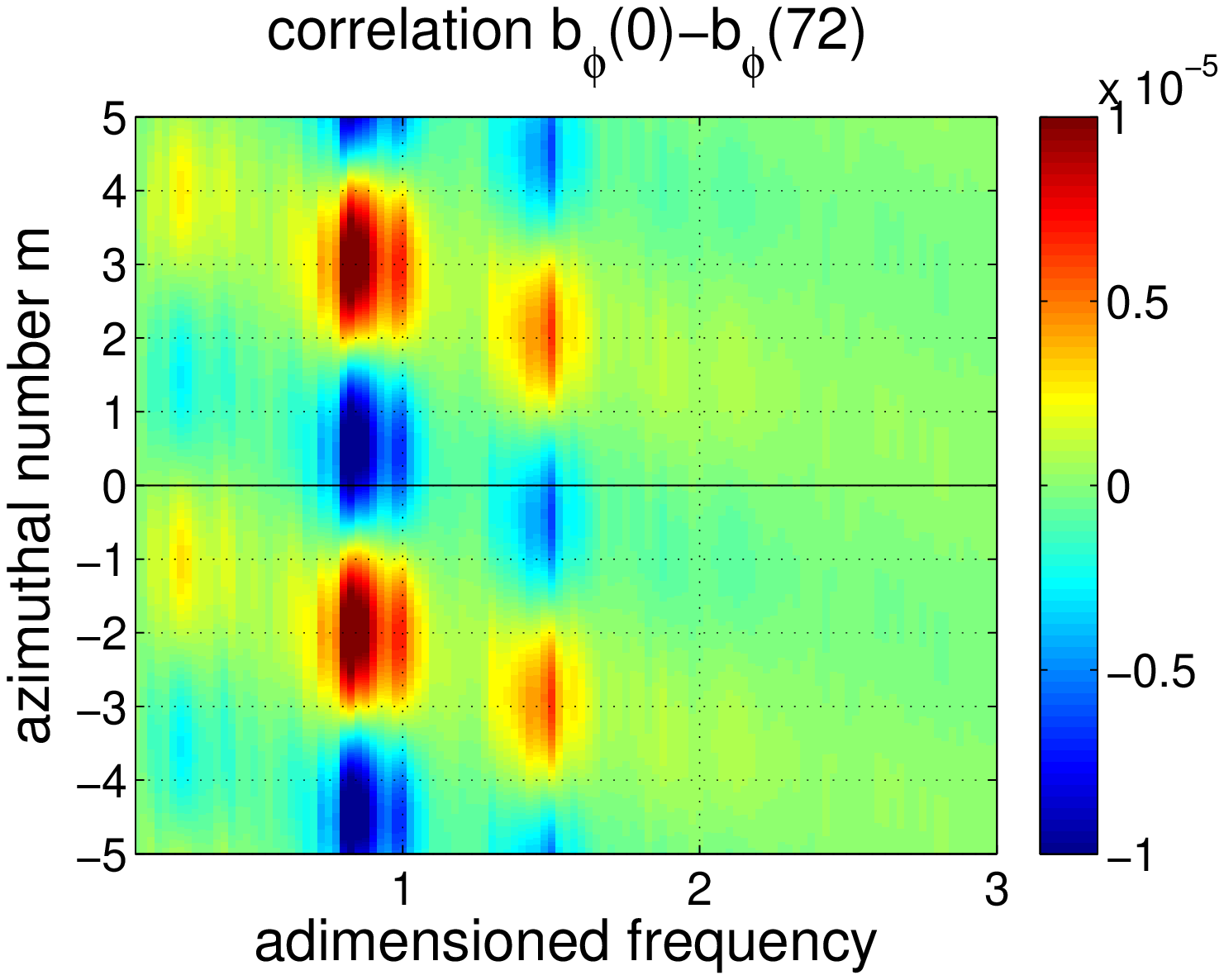}
	\end{center}
	\caption{Spectral bumps in our reference numerical simulation ($Pm=10^{-3}$, $\Rey=2\,611$ and $\Lambda=3.4 \times 10^{-2}$).
a) Frequency spectra of the three components of the magnetic field ($b_r$, $b_\theta$, $b_\varphi$ in this order going up) recorded at two different latitudes ($0^{\circ}$ and $41^{\circ}$ as indicated).
Same adimensionalization and plotting conventions as in figure \ref{fig:DTS_modes}a.
The time window used lasts 540 turns.
Note the spectral bumps and compare with figure \ref{fig:DTS_modes}a.
b) $m$ versus frequency plot for the same run, obtained from the covariance of two $b_\varphi$ time-series recorded at points $72^\circ$ apart in longitude (same latitude $=41^\circ$).
The red patches for negative $m$ indicate that the successive spectral bumps have a well-defined mode number $m$, whose absolute value increases with frequency (compare with figure \ref{fig:DTS_modes}b where the relevant $m$ are positive because the inner sphere spins in the negative direction).}
	\label{fig:spectra}
\end{figure*}

In order to compare the numerical results to the experimental measurements of the fluctuations, we perform a simulation over a long time-window (600 rotation periods) and record the magnetic field induced at the surface at selected latitudes.
We then compute the power spectra of these records as a function of frequency.
Typical spectra are shown in figure \ref{fig:spectra}a.
A sequence of bumps is clearly visible for both the radial and the azimuthal components of the magnetic field.
The spectra do not display power-law behaviour.


We note that long time series (longer than 300 rotation periods) are needed for the spectral bumps to show up clearly.
The bumps are not as pronounced as in figure{\ref{fig:DTS_modes}a}, but we note that $4 \,000$ turns were used for these experimental spectra.
It could also be that the bumps are enhanced at higher Reynolds number.

\subsection{Azimuthal mode number}\label{subsec:azimuthal_mode_number}


Pursuing further the comparison with the experimental results, we examine whether the various spectral bumps correspond to specific azimuthal mode numbers.
As in \citet{schmitt08}, we correlate the signals computed at the same latitude ($41^\circ$) but $72^\circ$ apart in longitude.
The signals are first narrow-band filtered, and we plot in figure \ref{fig:spectra}b the amplitude of the covariance (colour scale) as a function of the peak frequency of the filter, for time-delays between the two, converted into azimuthal mode number $m$ (y-axis).
As in the experiments (see figure \ref{fig:DTS_modes}b), we find that each spectral bump corresponds to a single dominant (here negative by convention) azimuthal mode number $m$.
The successive bumps have increasing azimuthal $m$ (1, 2 and 3).

\subsection{Full Fourier transform}\label{subsec:full_Fourier_transform}

\begin{figure*}
	\begin{center}
		\includegraphics[width=6.5cm]{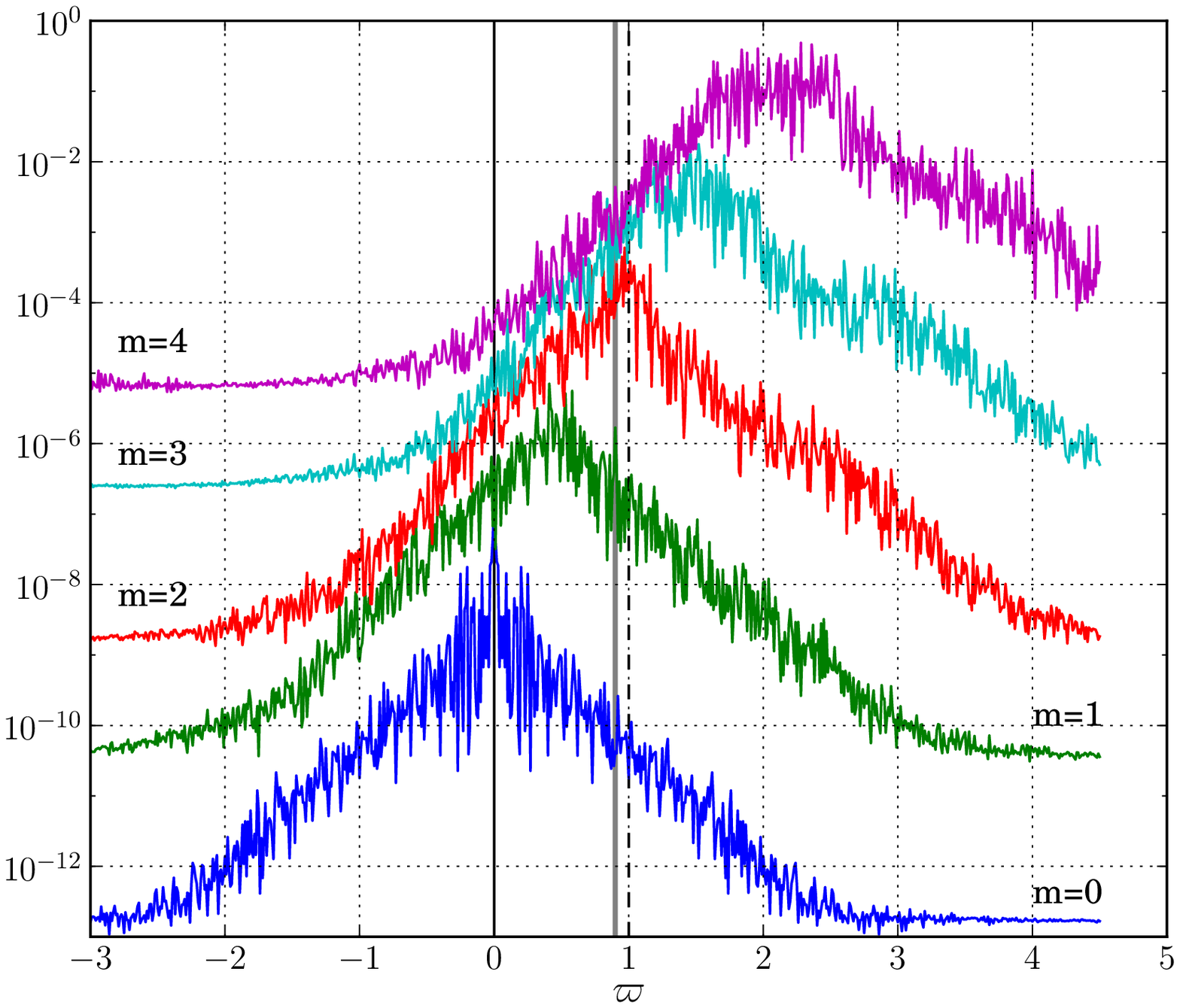}
		\includegraphics[width=6.5cm]{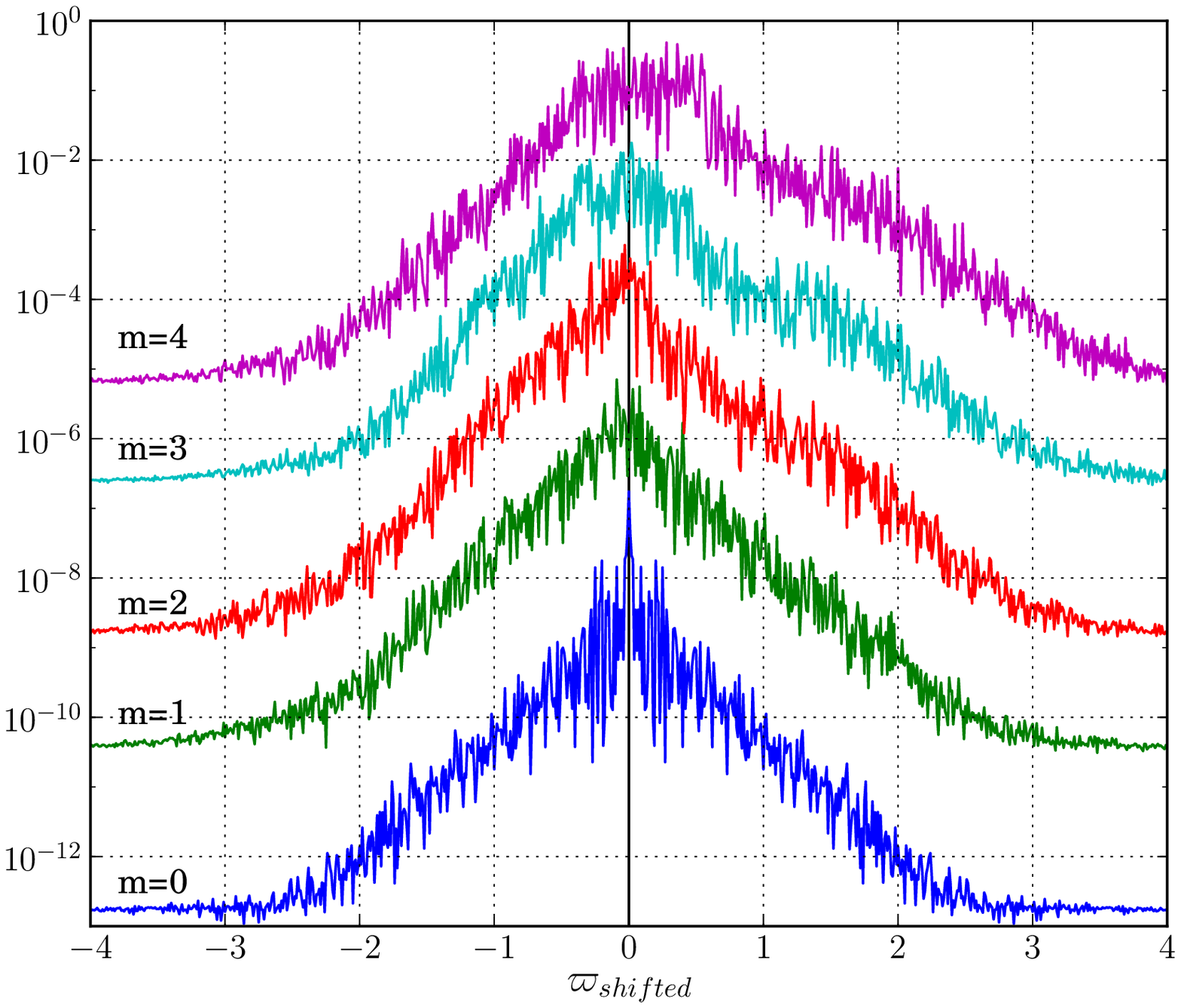}
	\end{center}
	\caption{Partial energy frequency spectra $\mathcal{E}_m(\varpi)$ of the equatorially-symmetric magnetic field at $r<0.55$.
Frequencies are normalized by $f$, the rotation frequency of the inner sphere (vertical dashed line).
a) Raw spectra for azimuthal mode numbers $m=0$ to $4$ (shifted vertically for clarity).
The vertical solid line indicates the frequency $\varpi^*$ at which we retrieve the $m=2$ mode structure in figure \ref{fig:fft_modes}.
b) Same as a) except that frequencies are shifted according to: $\varpi_{shifted} = \varpi - m f_{fluid}$ (see text).
}
	\label{fig:fft_spectra_1}
\end{figure*}

\begin{figure*}
	\begin{center}
		\includegraphics[width=6.5cm]{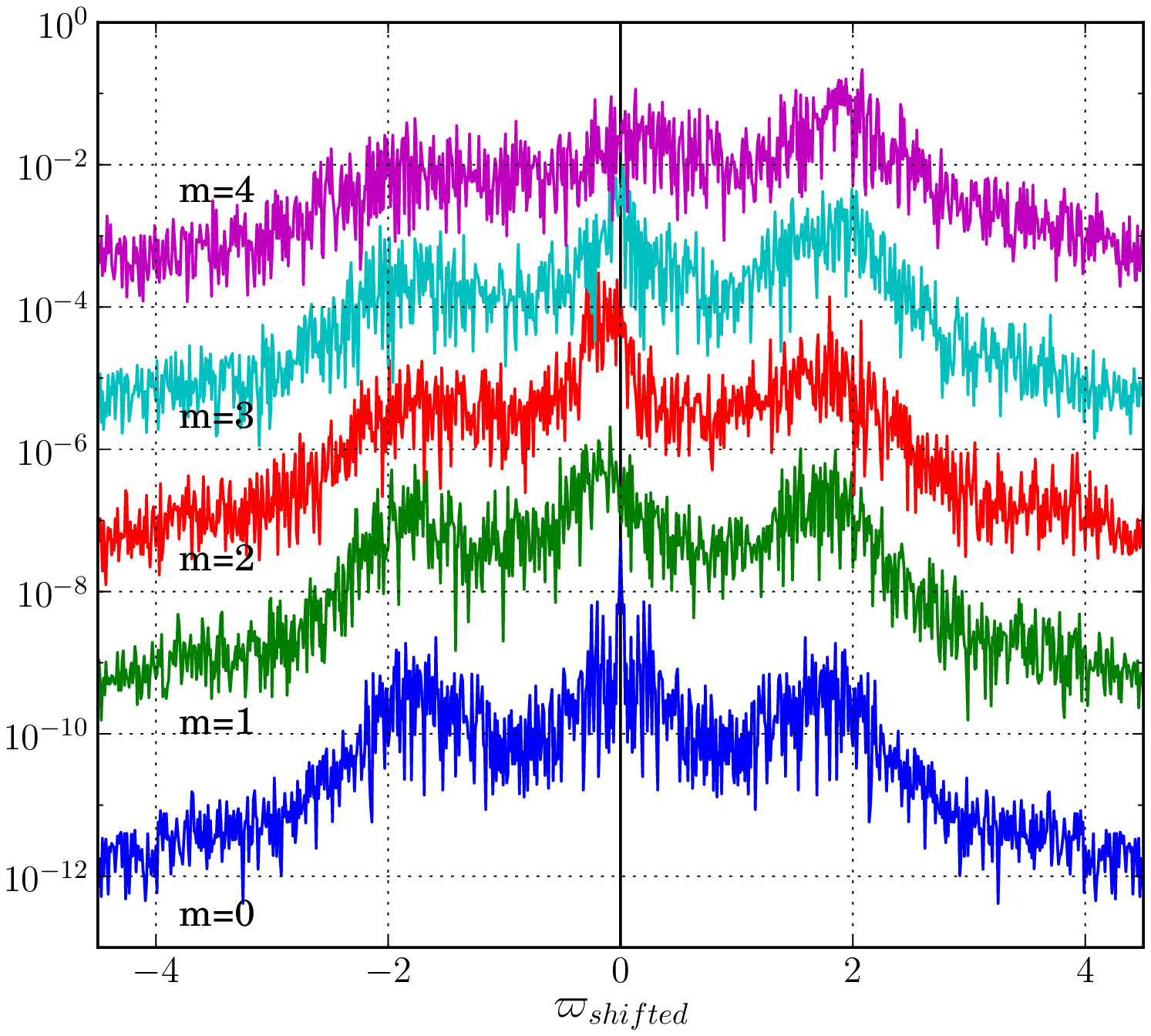}
		\includegraphics[width=6.5cm]{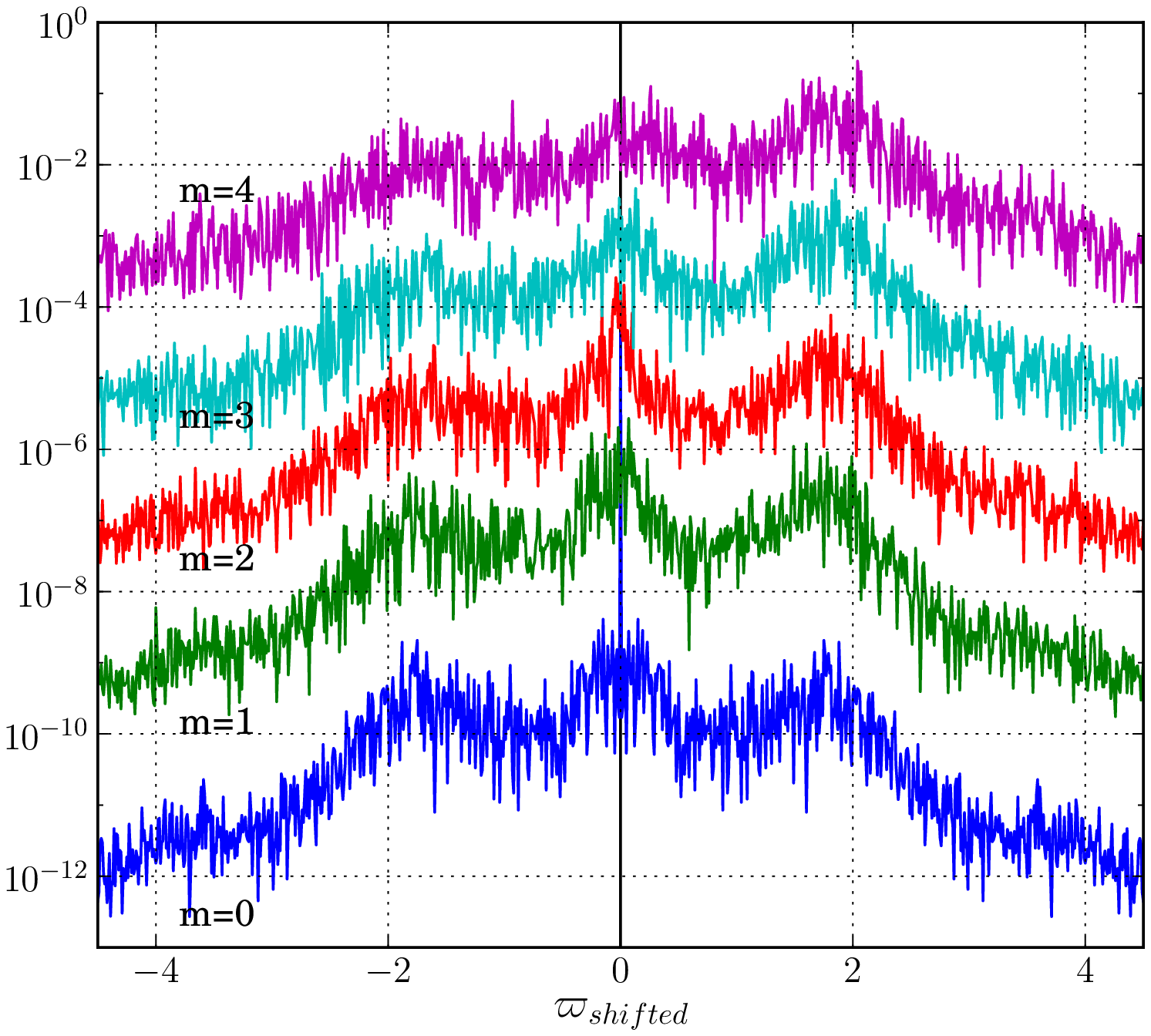}
	\end{center}
	\caption{Partial energy frequency spectra $\mathcal{E}_m(\varpi)$ of the magnetic field at the outer surface ($1<r<1.024$, $0.35 < \sin \theta < 0.6$) for different $m$. a) Equatorially symmetric part. b) Equatorially-antisymmetric part.
Frequencies are normalized by $f$, the rotation frequency of the inner sphere, and are shifted according to equation \ref{eq:shift}.
$Pm=10^{-3}$, $\Rey=2\,611$ and $\Lambda=3.4 \times 10^{-2}$.}
	\label{fig:fft_spectra_2}
\end{figure*}

In the numerical simulations, we can construct frequency spectra for each $m$.
When the stationary regime is reached, we record 900 snapshots of the full fields, regularly spaced in time during 100 rotation periods: $\boldsymbol{F}(r,\theta, \varphi, t)$, where $\boldsymbol{F}$ can be either $\boldsymbol{u}$ or $\boldsymbol{B}$.
A two dimensional Fourier transform in the azimuthal and temporal directions $\varphi$ and $t$ gives us a collection of complex vectors $\mathcal{\bold{F}}_m^\varpi$ representing the field for azimuthal number $m$ and discrete frequency $\varpi$, such that
\begin{equation}
	\boldsymbol{F}(r,\theta,\varphi,t) = \sum_m \sum_\varpi \mathcal{\bold{F}}_m^\varpi(r,\theta) e^{i(m\varphi - \varpi t)}.
	\label{eq:fft}
\end{equation}

Note that the sign of the frequency has thus a precise meaning: positive (negative) frequencies correspond to prograde (retrograde) waves or modes.

This allows us to compute partial energy spectra
\begin{equation}
\mathcal{E}_m(\varpi) = \int_{r_1}^{r_2} \int_{\theta_1}^{\theta_2} \left| \left| \mathcal{\bold{F}}_m^\varpi(r,\theta) \right| \right|^2 \, r \, \sin\theta \, d\theta \, dr.
	\label{eq:fft_energy}
\end{equation}

Magnetic partial energy spectra $\mathcal{E}_m(\varpi)$ for the inner region ($0.35<r<0.55$, $0<\theta<\pi$) are shown in figure \ref{fig:fft_spectra_1}a for  $m=$ 0 to 4.
They are dominated by a single peak, which moves towards positive (prograde) frequencies as $m$ increases.
This can be interpreted as the advection of stationary or low-frequency structures by a prograde fluid velocity.
We therefore shift the frequency of the spectra in figure \ref{fig:fft_spectra_1}b, according to:

\begin{equation}
\varpi_{shifted} = \varpi - m f_{fluid}
	\label{eq:shift}
\end{equation}
where $f_{fluid} = 0.5$.
We choose this value because it provides a good alignment of the spectral peaks and is compatible with the bulk fluid velocity in the outer region beneath the boundary layer (see figure \ref{fig:u_field_uw_upol_merid_Re_2600}a).
This shift explains the linear evolution of the frequencies of the spectral bumps with $m$ observed both in the $DTS$ experiment (figure \ref{fig:DTS_modes}b) and in the simulation (figure \ref{fig:spectra}b).
It means that the peaks are caused by the advection of periodic structures by the mean flow, or by a non-dispersive wave.

We now turn to the partial energy spectra of the magnetic field at high latitude ($0.35 < \sin \theta < 0.6$), at the surface of the sphere ($1<r<1.024$), displayed in figure \ref{fig:fft_spectra_2}a and \ref{fig:fft_spectra_2}b for the equatorially symmetric and anti-symmetric parts, respectively.
The frequencies are again shifted according to equation \ref{eq:shift}.
This time, three peaks dominate the $m=0$ spectra.
The spectrum is symmetric with respect to $\varpi= 0$ since there cannot be prograde or retrograde propagation for $m=0$: only latitudinal propagation or time-oscillations are permitted.
The lateral peaks yield a frequency $\varpi^{\dagger} \simeq 1.8$.
As $m$ increases, the lateral peak becomes dominant in the prograde direction, while it vanishes in the retrograde direction.
We note that the peaks are well aligned in these shifted representation, meaning that these secondary fluctuations are also advected at roughly the same angular velocity as the central peak.
But both stationary and propagating waves are required to explain that this peak is not at zero frequency, and that it has both a prograde and a retrograde signature, and that the former dominates for $m \neq 0$.

Note that these secondary peaks do not show up in the regular frequency spectra or $m$-plots of point-measurements (figure \ref{fig:spectra}).
This illustrates the interest and potential of the full Fourier transform method that we developed.

\subsection{Mode structure}\label{subsec:mode_structure}

\begin{figure*}
	\begin{center}
		\includegraphics[width=6.5cm]{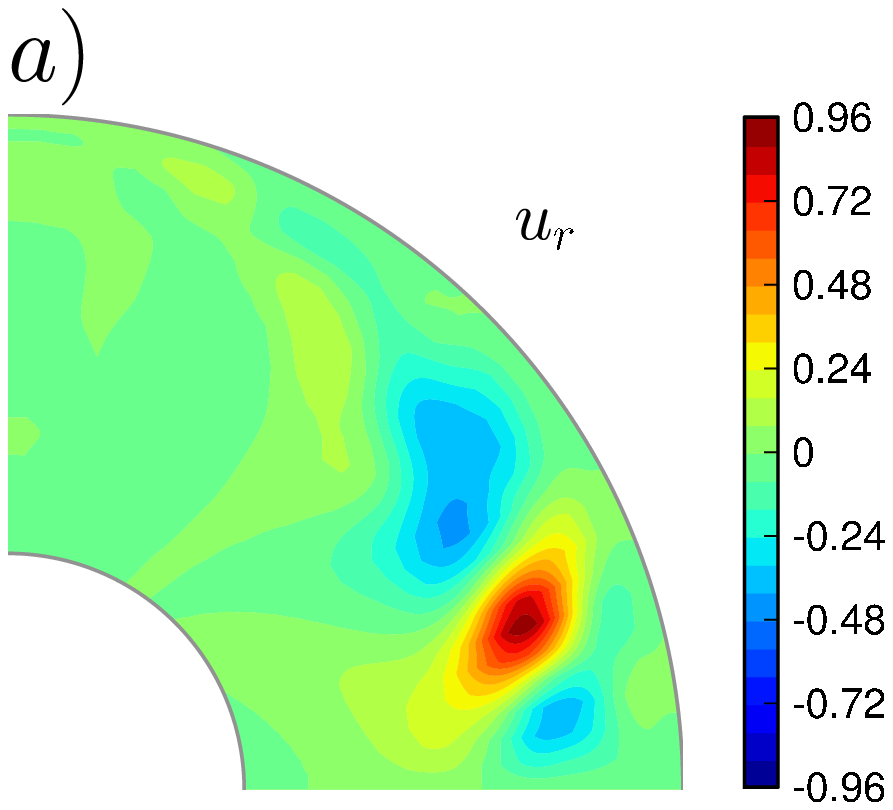}
		\includegraphics[width=6.5cm]{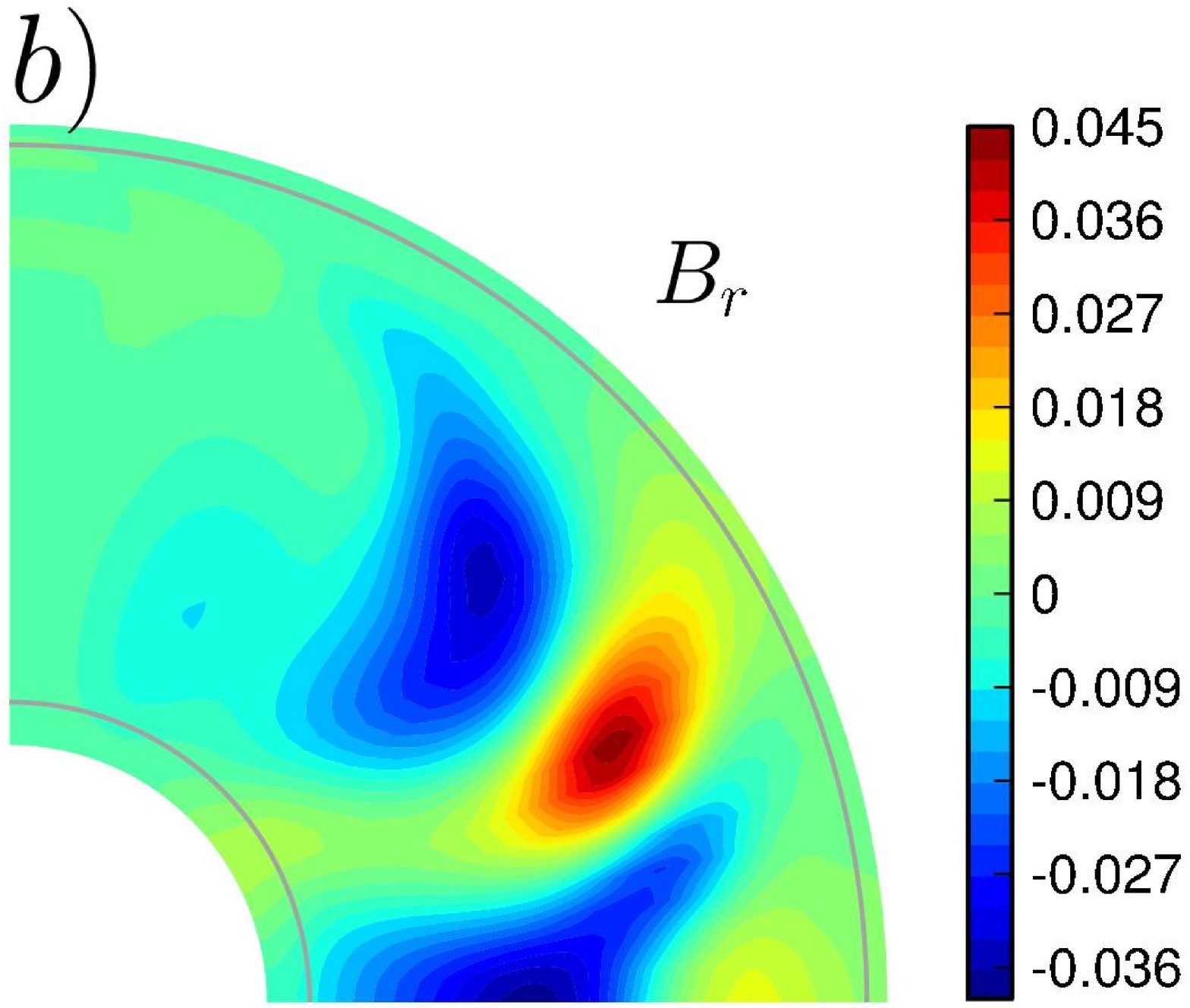}\\
		\includegraphics[width=6.5cm]{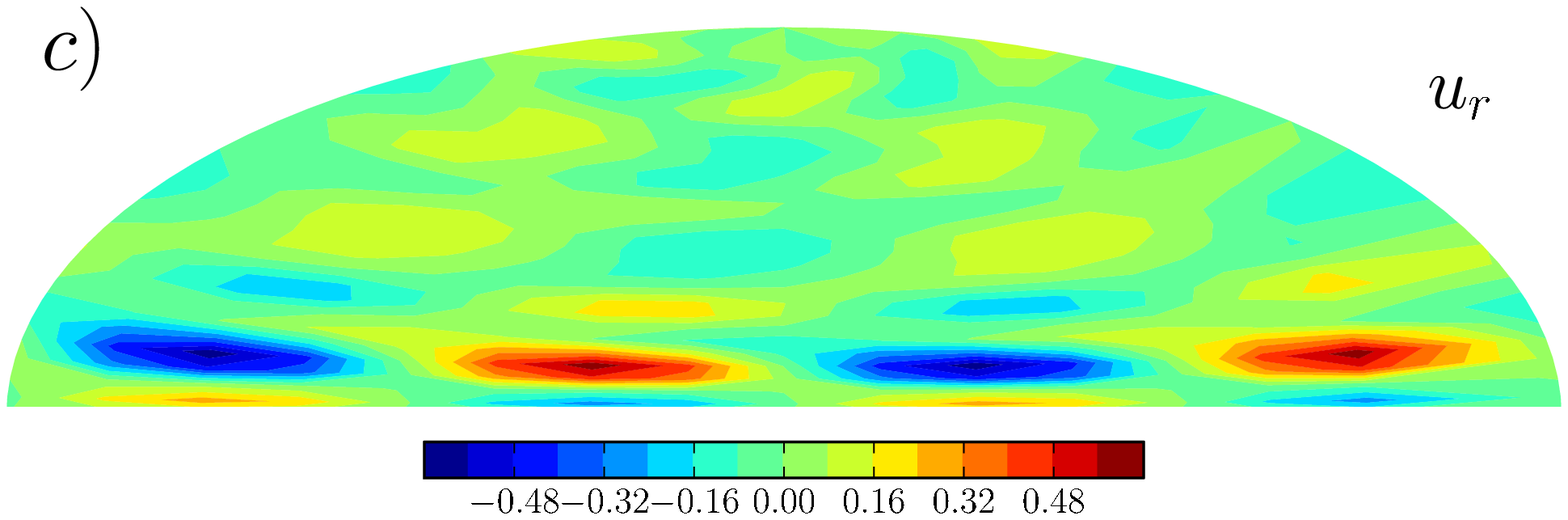}
		\includegraphics[width=6.5cm]{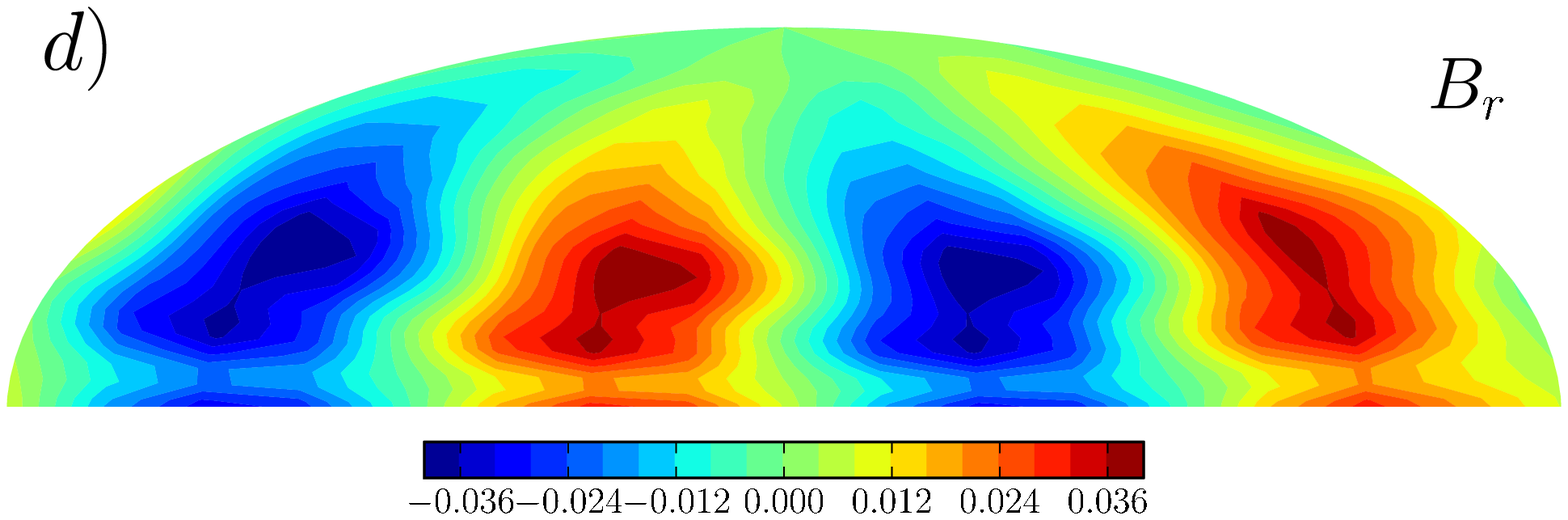}		
	\end{center}
	\caption{$m=2$ mode structure given by the full Fourier transform method, for frequency $\varpi^* = 0.9$, which corresponds to the $m=2$ peak in figure \ref{fig:fft_spectra_1}a. $Pm=10^{-3}$, $\Rey=2\,611$ and $\Lambda=3.4 \times 10^{-2}$. Energy densities are normalized by $E_K^{0}=\frac{1}{2} \rho \Omega^2 r_i^2$. a) and b) isovalues in the $\varphi=0^{\circ}$ meridional plane of the radial velocity and radial magnetic field, respectively. c) and d) mapview of the isovalues at $r=0.95$ of the radial velocity and radial magnetic field, respectively.}
	\label{fig:fft_modes}
\end{figure*}

Picking the frequency that yields the maximum spectral energy density for a given $m$, we derive the structure of the corresponding mode. One example for $m=2$ is shown in figure \ref{fig:fft_modes}, where we plot, for both $u_r$ and $B_r$, the structure of the mode in a meridional plane  and in map view at $r=0.95$. We selected a mode for which $B_r$ is symmetric with respect to the equator (and thus $u_r$ is anti-symmetric).

The meridional map for $u_r$ reveals structures in the outer region, while $B_r$ shows similar patterns that extend deeper down to the inner sphere.
While the map view of $u_r$ near the outer boundary displays short wavelength structures, we find it remarkable that the structure of $B_r$ is very smooth and very similar to those retrieved in the $DTS$ experiment, and well-accounted for in the linear modal approach of \citet{schmitt12}.

 
%

\section{Fluctuations and instabilities}\label{sec:fluctuations}

Having shown that our numerical simulations recover the essential features of the modes and spectra of the $DTS$ experiment, even though their Reynolds number is much smaller, we now examine where and how the modes are excited.
The first guide we use is the location of the largest fluctuations.

\subsection{Energy fluctuations}\label{subsec:energy_fluctuations}

We compute the kinetic energy density as $\delta E_K=\frac{1}{2} \rho \langle (u - \langle u \rangle)^2\rangle$ and the magnetic energy density as $\delta E_M=\frac{1}{2\mu_0} \langle ( b - \langle b \rangle)^2 \rangle$, where $\langle \rangle$ denotes time-averaging.
We normalize both by a reference kinetic energy density $E_K^{0}=\frac{1}{2} \rho \Omega^2 r_i^2$, and we integrate over azimuth.

\begin{figure*}
	\begin{center}
		\includegraphics[width=6.5cm]{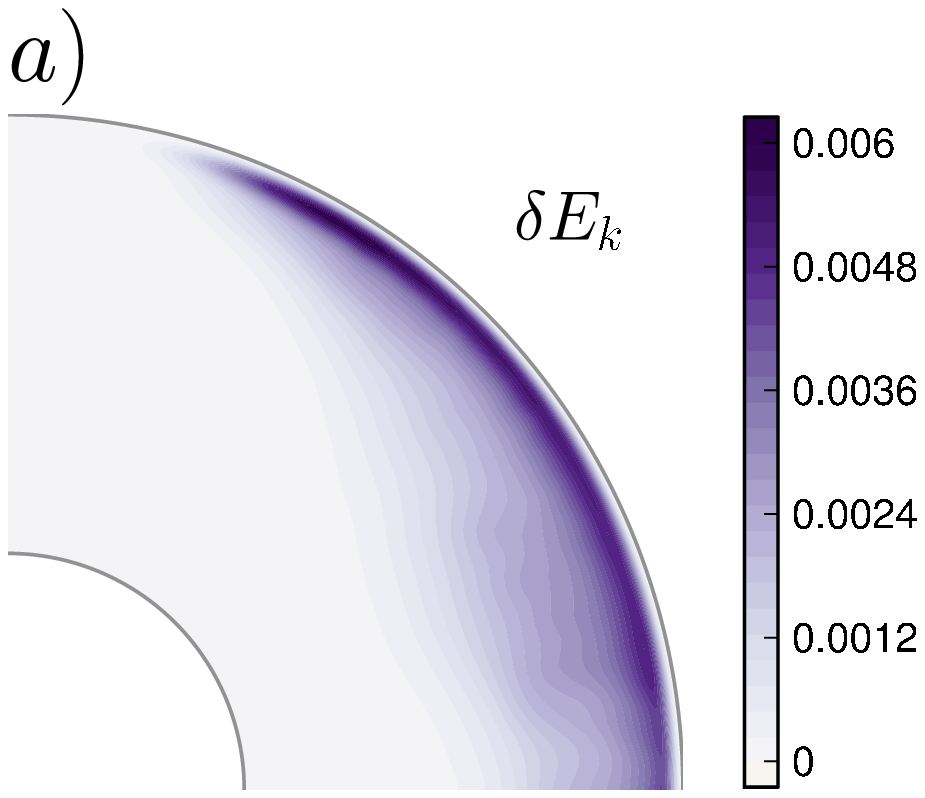}
		\includegraphics[width=6.5cm]{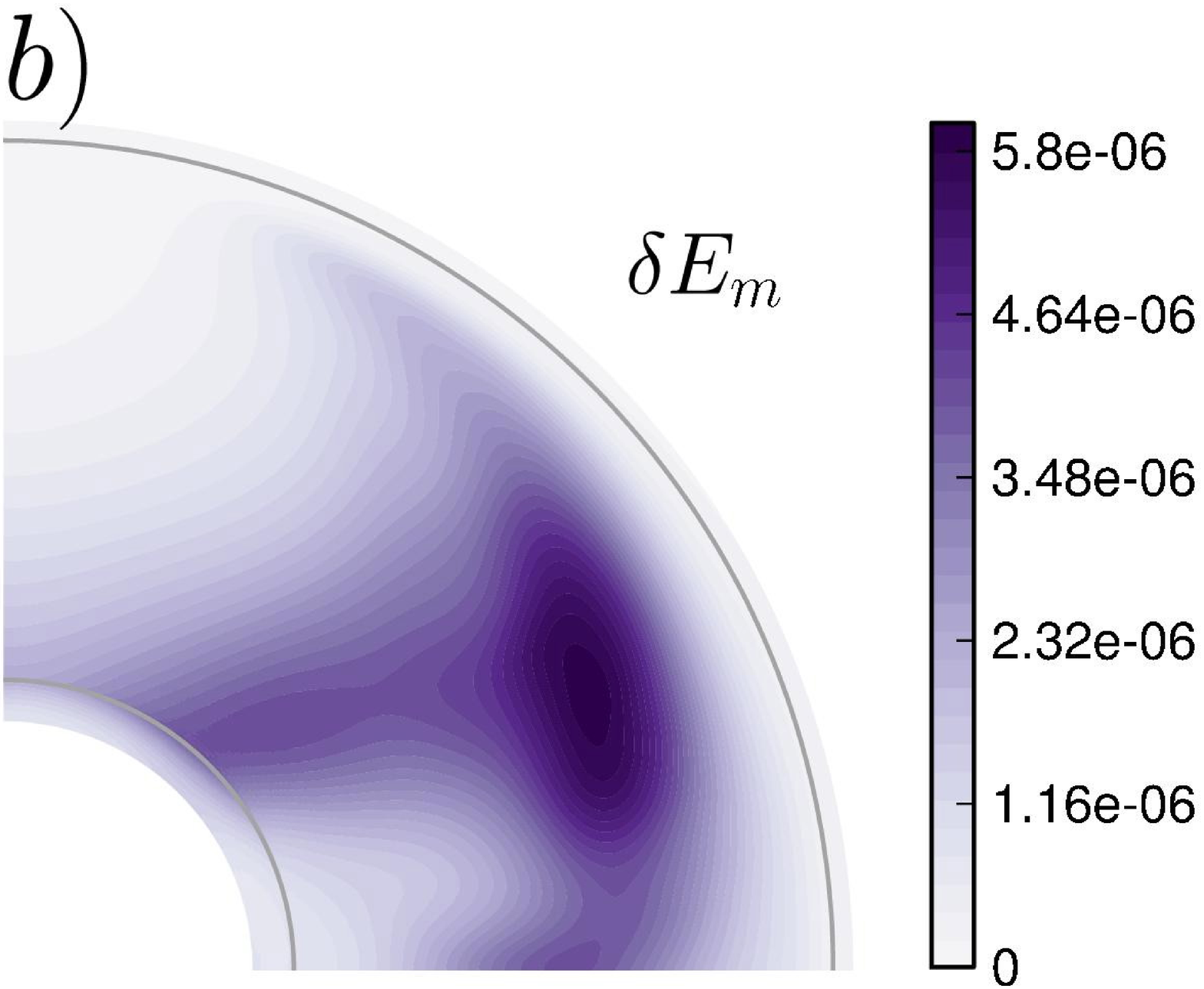}
	\end{center}
	\caption{Time-and-azimuth-averaged meridional map of the energy density of the fluctuations. a) Kinetic energy. b) Magnetic energy. $Pm=10^{-3}$, $\Rey=2\,611$ and $\Lambda=3.4 \times 10^{-2}$. The energy densities are normalized by $E_K^0$.}
	\label{fig:delta_energy_u_f_r_theta_Re_2600}
\end{figure*}

Figure \ref{fig:delta_energy_u_f_r_theta_Re_2600} displays the resulting kinetic and magnetic energy densities of the fluctuations in a meridional plane for our reference simulation.
We observe that the kinetic energy is maximum in the outer boundary layer, while the (much weaker) magnetic energy extends all the way to the inner sphere.

\begin{figure*}
	\begin{center}
		\includegraphics[width=6.5cm]{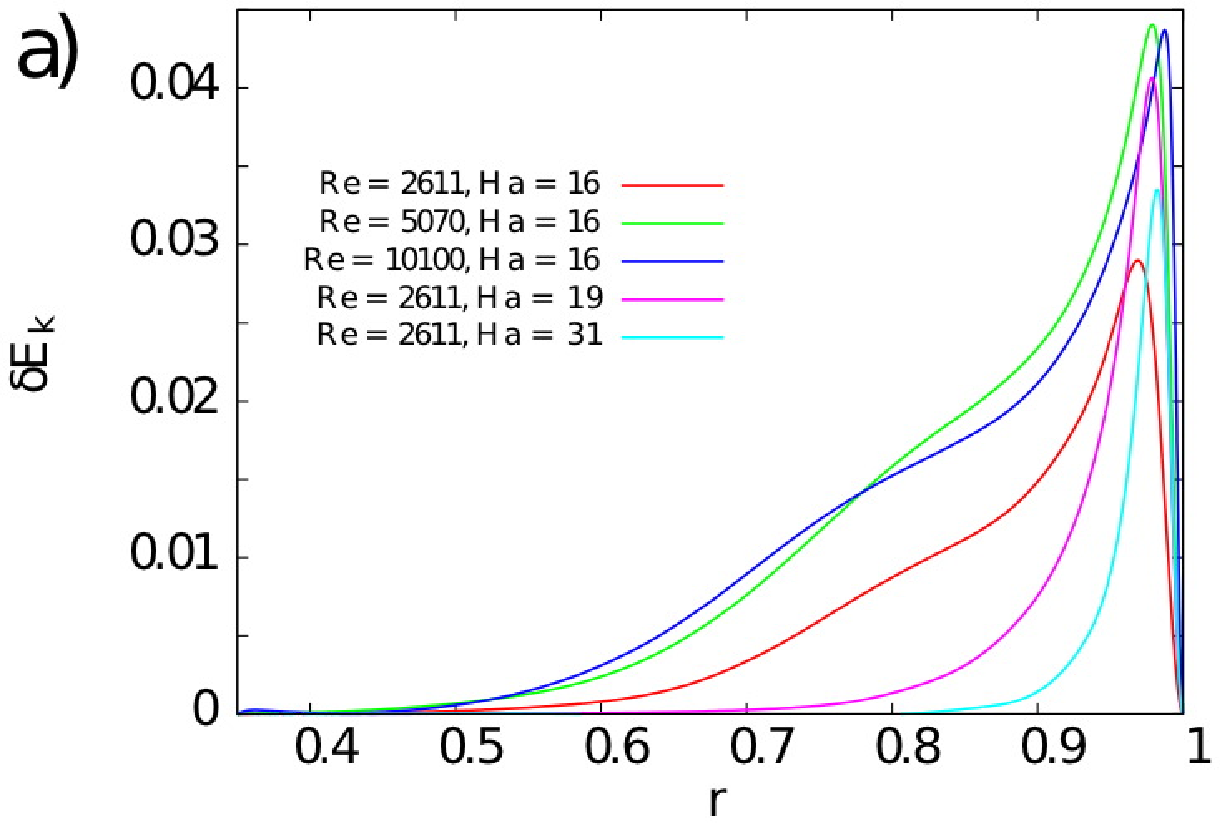}
		\includegraphics[width=6.5cm]{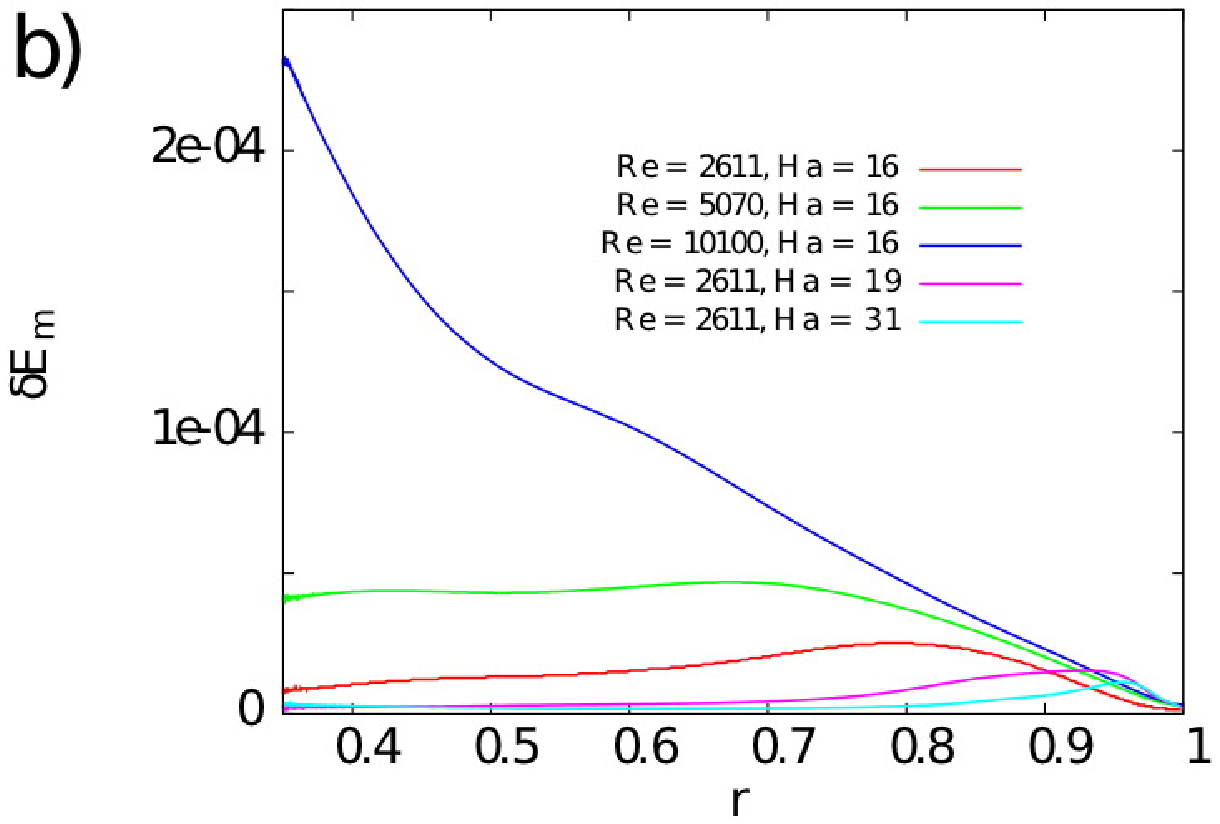}
	\end{center}
	\caption{Time-and-colatitude-averaged energy density of the fluctuations as a function of radius $r$. a) Kinetic energy. b) Magnetic energy.
The energy densities are normalized by $E_K^0$. 
The plots compare the fluctuations obtained for three simulations with the same Hartmann number ($Ha=16$) and increasing Reynolds numbers ($\Rey=2\,611$, $5\,070$ and $10\,100$), and three simulations with the same Reynolds number ($\Rey=2\,611$) and  increasing Hartmann numbers ($Ha=16$, $19$ and $31$).
Other dimensionless numbers as in Table \ref{tab:numbers}.
Note that the magnetic energy is smaller than the kinetic energy by three orders of magnitude, but increases strongly toward the inner sphere for the simulation with the highest Reynolds number.
}
	\label{fig:comp_delta_energy_u_b_Res_Has}
\end{figure*}

Figure \ref{fig:comp_delta_energy_u_b_Res_Has} shows the radial profiles obtained after integration over the colatitude $\theta$.
It illustrates the effect of varying the Reynolds and the Hartmann numbers of the simulations.
The fluctuations remain strongest in the outer boundary layer, but extend deeper inside the fluid with increasing Reynolds number.
For the highest Reynolds number ($\Rey = 10\,100$), this causes the magnetic energy to strongly increase with depth, as the fluctuations interact with the larger imposed magnetic field near the inner sphere.

\subsection{Origin of the fluctuations}\label{subsec:origin_of_fluctuations}
It is beyond the scope of this paper to characterize the complete scenario by which instabilities develop in our geometry.
However, we find it important to identify where the instabilities originate in order to understand the excitation of the modes we observe and extrapolate to other situations.
The energy maps (figure \ref{fig:delta_energy_u_f_r_theta_Re_2600}) as well as movies of the simulations (see online supplementary material) strongly suggest that fluctuations are initiated in the outer boundary layer.
There is a large azimuthal velocity drop across the outer boundary layer, from the vigorously entrained fluid in the core flow to the outer container at rest.

A detailed inspection of the numerical simulations reveals two types of instabilities, which do not occur in the same region but appear to be coupled through the meridional circulation: i) instability of a B\"odewadt type layer at high latitude; ii) secondary instability of a centripetal jet at the equator.
We use the meridional snap-shots of figure \ref{fig:added} to illustrate these two mechanisms.

\subsubsection{B\"odewadt layer instability}
\begin{figure*}
	\begin{center}
		\includegraphics[width=4.3cm]{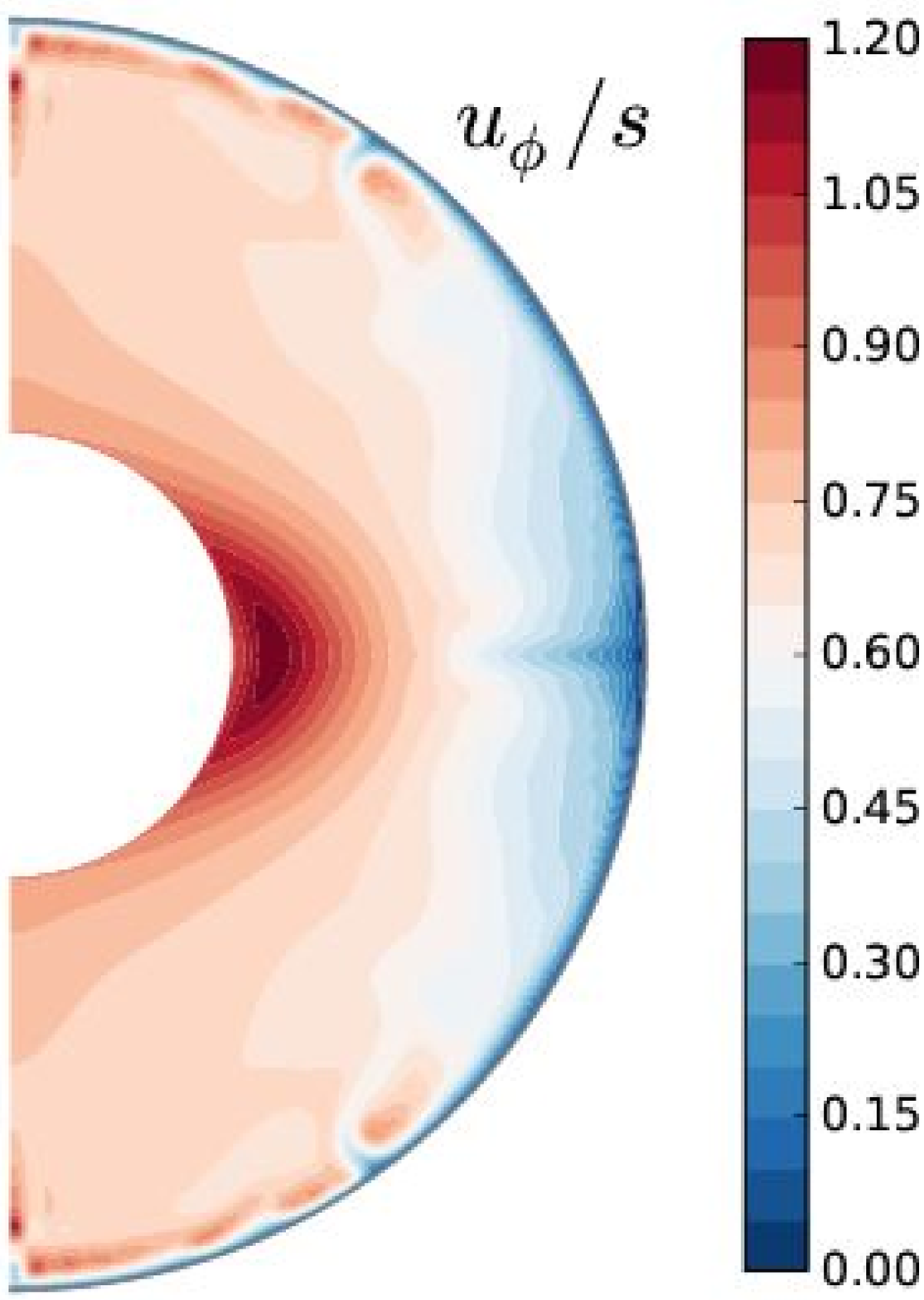}
		\includegraphics[width=4.3cm]{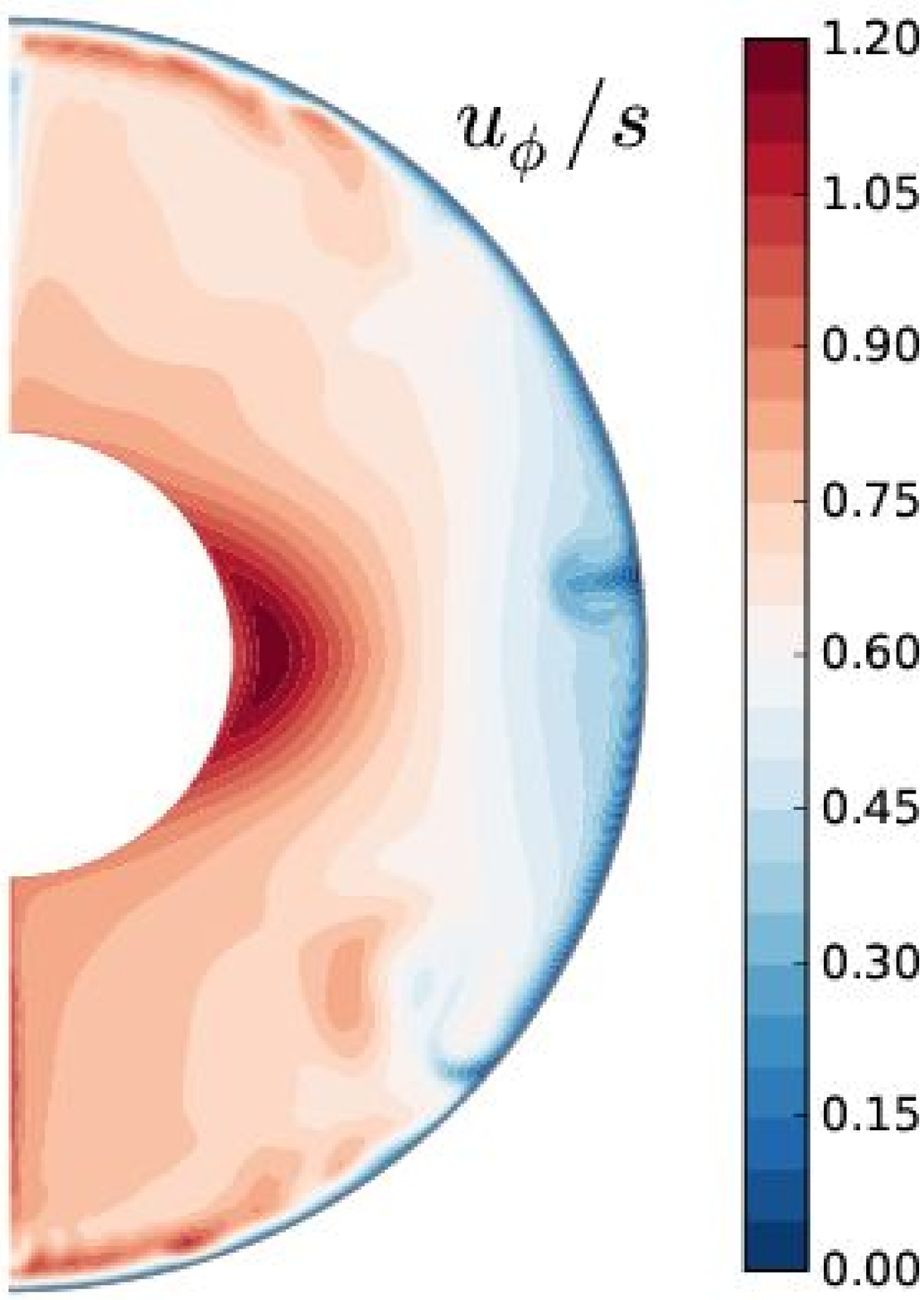}
		\includegraphics[width=4.3cm]{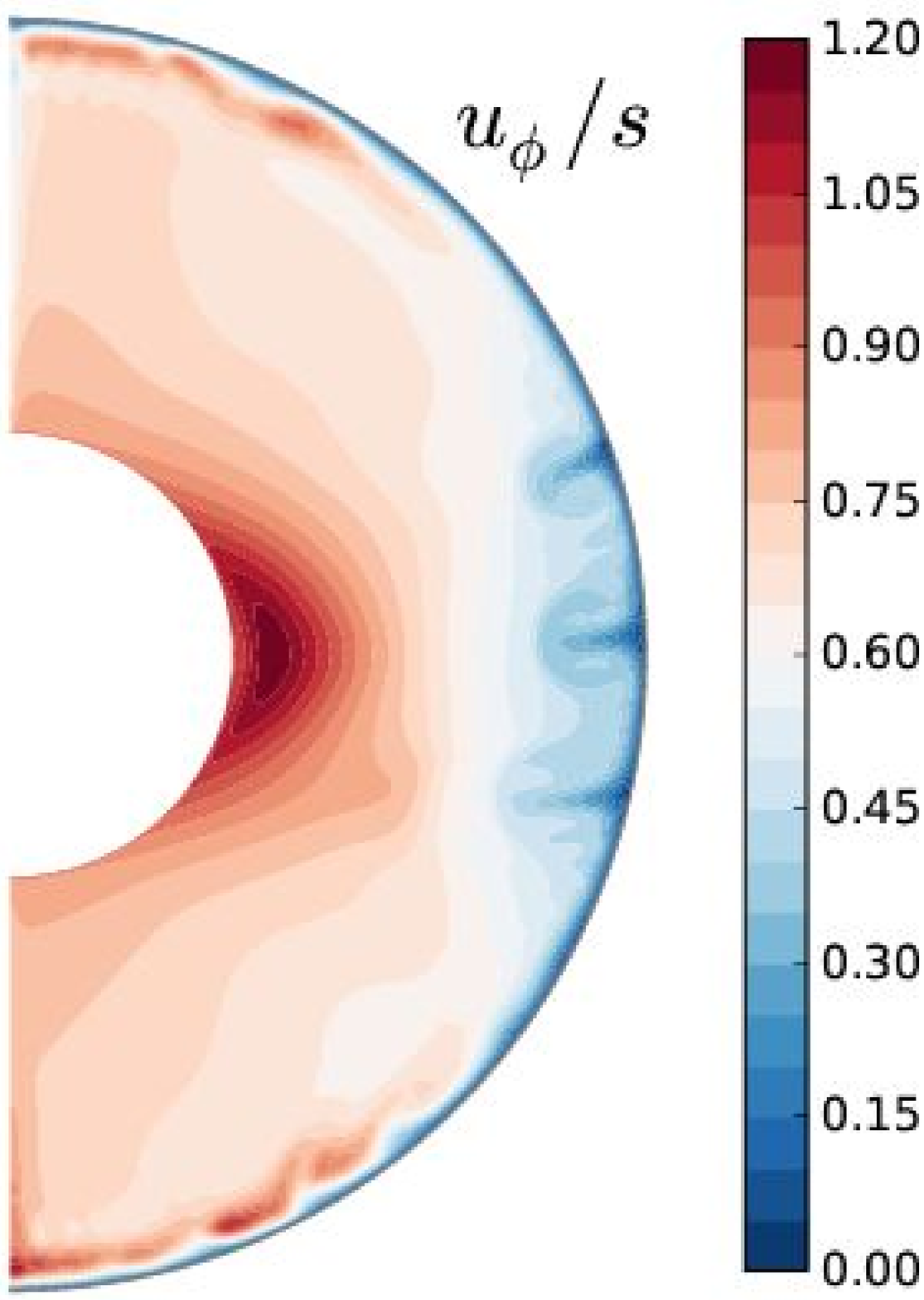}
	\end{center}
	\caption{Meridional snap-shots of the angular velocity of the fluid at successive times $t$ (given in number of turns of the inner sphere) in our reference simulation. $Pm=10^{-3}$, $\Rey=2\,611$ and $\Lambda=3.4 \times 10^{-2}$. a) $t=17.5$; b) $t=93.8$; c) $t=675$.
}
	\label{fig:added}
\end{figure*}

The flow that appears when a fluid rotates at constant angular velocity above a flat disk at rest has been studied by  \citet{Bodewadt40} who has found the analytic expression of the boundary layer that develops at the surface of the disk.
It is characterized by a large overshoot in the azimuthal velocity profile caused by the centripetal radial circulation.
As shown by \citet{Lingwood97}, this boundary layer is particularly unstable, and several teams have analyzed the instabilities that take place \citep[e.g.,][]{Savas87, Lingwood97, Lopez09, Gauthier99, Schouveiler01}.
Two types of instabilities have been reported: axisymmetric rolls that propagate inwards (following the centripetal circulation of the boundary layer), and spiral rolls.
A B\"odewadt-type situation is encountered in our geometry at high latitude.
Figure \ref{fig:u_field_uw_upol_merid_Re_2600}a shows a clear overshoot of the angular velocity at latitudes above about $40^\circ$, linked to a polewards meridional circulation.
In this region, we observe polewards propagating axisymmetric rolls in our simulations when the inner sphere is spun from rest.
This is best seen in the movies provided as supplementary material online, but the signature of the rolls is clearly visible at high latitude in the three snap-shots of figure \ref{fig:added}.

It also shows up in the spatio-temporal representations of the instabilities at $r=0.95$, just beneath the outer boundary layer (figure \ref{fig:spatio_temporal}).
Figure \ref{fig:spatio_temporal}a gives the axisymmetric part of $u_{\theta}$ as a function of time ($x$-axis) and latitude ($y$-axis) from $-\pi / 2$ to $\pi / 2$.
The high latitude rolls show up as successive inclined lines in this time-latitude plot.

Note that similar high latitude rolls have been reported in axisymmetric simulations of magnetized spherical Couette flow \citep{hollerbach07, Brito11}.

\subsubsection{equatorial centripetal jet instability}
\begin{figure*}
	\begin{center}
		\includegraphics[width=11cm]{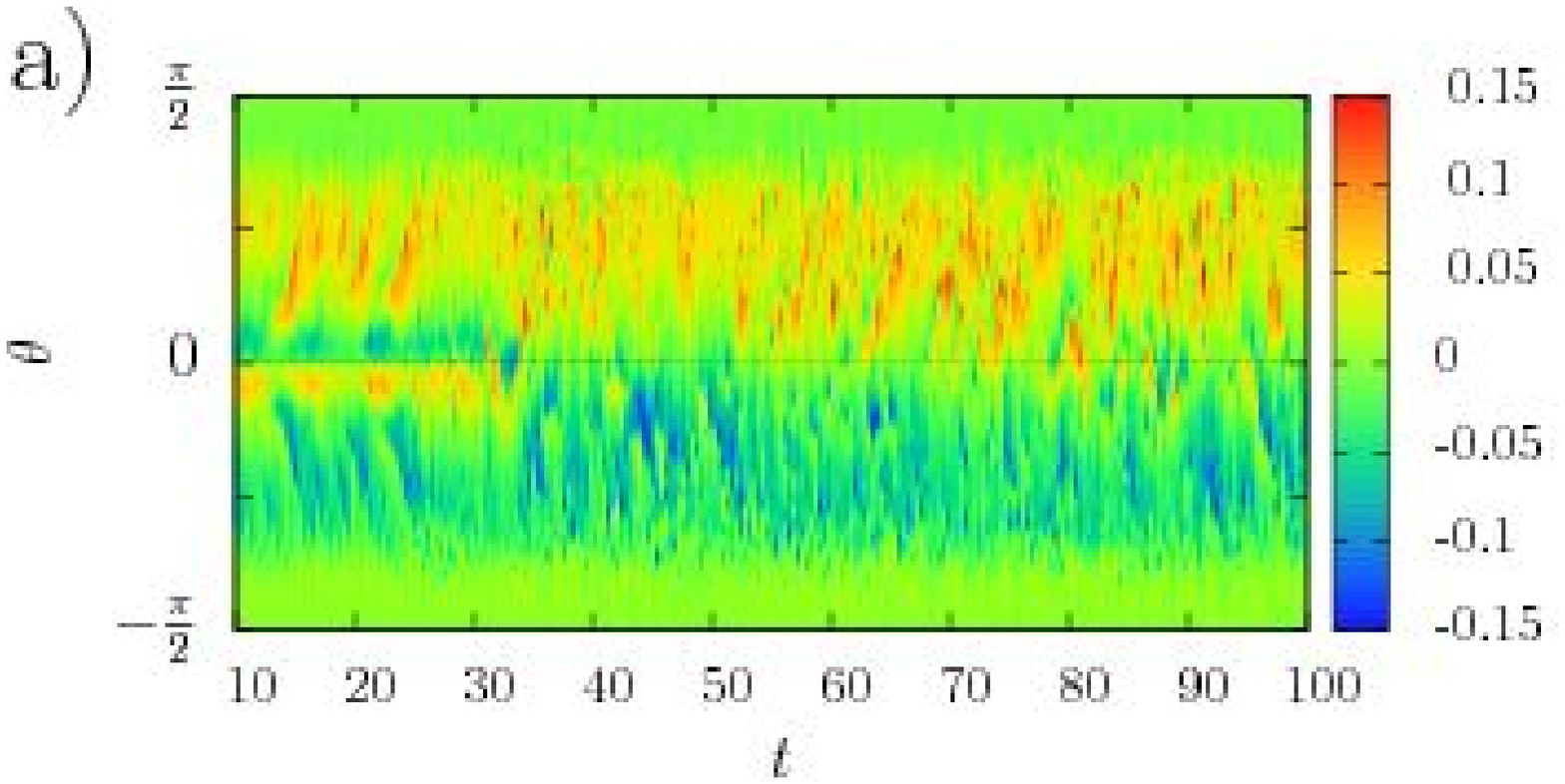}\\
		\includegraphics[width=11cm]{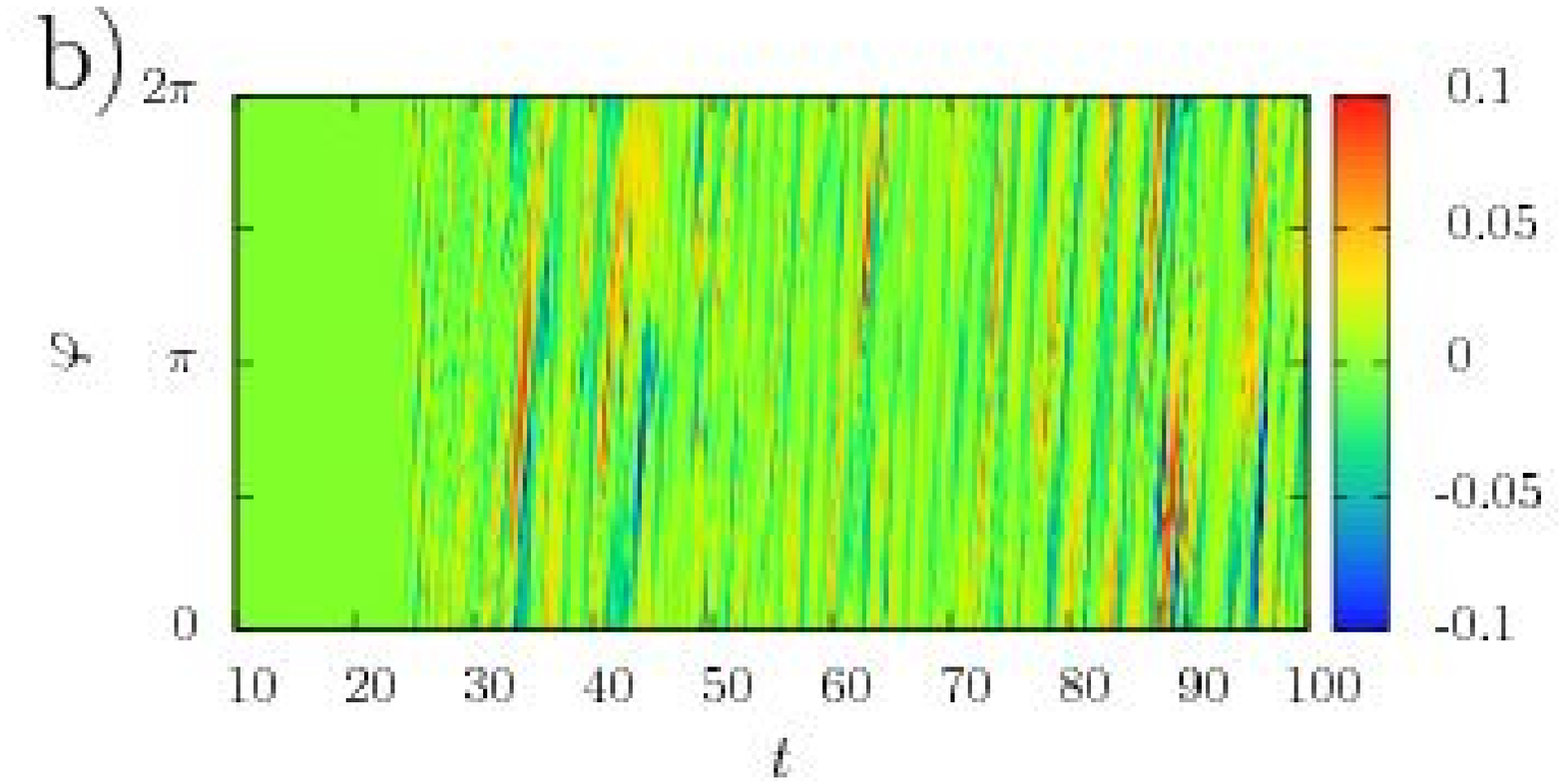}
	\end{center}
	\caption{Spatio-temporal representation of the velocity fluctuations in the simulation at radius $r=0.95$. $Pm=10^{-3}$, $\Rey=2\,611$ and $\Lambda=3.4 \times 10^{-2}$.
a) $u_\theta$ isovalues of the axisymmetric flow ($m=0$) in a latitude-time plot.
b) $u_\varphi$ isovalues at $10^\circ$ latitude in a longitude-time plot ($m=0$ term removed).
Velocities are normalized by the reference velocity $\Omega r_i$.
Time is given in rotation periods.
}
	\label{fig:spatio_temporal}
\end{figure*}

A different kind of instability takes place at the equator.
In figure \ref{fig:spatio_temporal}a, polewards migrating rings yield a butterfly pattern, which also reveals that the equatorial downwelling instability creates a meridional circulation of opposite sign around the equator for $t<30$ turns.
\citet{Brito11} report this equatorial counter-rotating cell for some parameters in their axisymmetric equatorially-symmetric simulations. 
As shown in figure \ref{fig:added}a, it is associated to a sheet that draws fluid -and reduced angular momentum- from the outer boundary inside the sphere, in the equatorial place.
It can probably be described as a centrifugal Taylor-G\"ortler instability \citep{saric94}, similar to those observed by \citet{Noir09} in libration-induced flows in a sphere.
As usually happens for these vortices, the non-linear evolution of the instability leads to mushroom-type downwellings (figures \ref{fig:added}b,c and online movies (supplementary material)).
 
At time $t \simeq 24$ turns, both the equatorial symmetry and the axisymmetry are broken by an $m=3$ undulation, which rapidly disrupts the pattern of the fluctuations.
Note however that axisymmetric bursts persist throughout, with amplitudes comparable to the initial ones.
They propagate mostly polewards, but some occasionally cross the equator.

The $m=3$ undulation is best observed in the spatio-temporal plot of figure \ref{fig:spatio_temporal}b, which displays the non-axisymmetric fluctuations of the azimuthal velocity at a latitude of $10^{\circ}$, as a function of time and longitude ($y$-axis) from $0$ to $2\pi$.
Until $t \simeq 24$ turns, there is no non-axisymmetric fluctuation, but at $t \simeq 24$ turns an $m=3$ mode appears (there are three maxima on a vertical line for a given $t$).
After a few turns, this initial $m=3$ undulation is replaced by chaotic fluctuations with dominant $m=1$ and $m=2$ contributions, which travel in the prograde direction with approximately the same velocity (given by the slope of the color streaks in this figure).

The $m=3$ secondary instability is similar to those observed in non-magnetic spherical Couette \citep{dumas91, guervilly10} or with an axial magnetic field \citep{hollerbach09}.
In these cases, it takes place on the centrifugal equatorial jet, which is a primary feature of these flows.

In our case note that, while the equatorial counter-rotating cell is essential for the centripetal jet to form, the time-averaged meridional circulation (shown in figure \ref{fig:u_field_uw_upol_merid_Re_2600}b) does not show this feature, as if the interplay of the developed instabilities had erased it.

\subsubsection{threshold of instability}
Although we do not intend to decipher the complete scenario of instability, we have determined the threshold of instability, which is found at $\Rey_c = 1 \, 860$, with a critical azimuthal mode number of $2$.
Interestingly, it seems that the two (coupled) instabilities described above are present from this threshold.
We can relate this threshold to the critical Reynolds number of the boundary layer.
Following \citet{Lingwood97}, we define the local Reynolds number $re = \omega^\star s^\star l/\nu$, where $l$ is the thickness of the laminar boundary layer: $l = \sqrt{\nu/\omega^\star}$, with $\omega^\star$ the dimensional angular velocity of the fluid with respect to the wall at a position specified by its dimensional cylindrical radius $s^\star$.
We can relate it to our global Reynolds number $\Rey$ by:
\begin{equation}
	re = \omega s \sqrt{\frac{r_o}{r_i}} \sqrt{\Rey},
\end{equation}
where the angular velocity $\omega$ just outside the boundary layer is adimensioned by $2 \pi f$ and the cylindrical radius $s$ by $r_o$, as before.
Picking $\omega \simeq 0.7$ at a latitude of $45^\circ$ ($s \simeq 0.7$) from figure \ref{fig:mean_omega_profiles}, we get $re_c \simeq 36$.
This is somewhat larger than the critical value of $21.6$ found by \citet{Lingwood97} for the absolute instability of a pure B\"odewadt layer, for which she predicts a critical mode number $m_c = \beta_c re_c = -0.1174 \times 21.6 \simeq -2.5$, which is not incompatible with our observation of an initial $m=2$ or $m=3$ pattern.
It is difficult to assess whether the higher threshold we get is due to a stabilizing effect of the magnetic field as in \citet{Moresco04}, or to the spherical geometry.

In any case, all the experiments analyzed by \citet{schmitt08, schmitt12} are far above this threshold.
Our main conclusion at this stage is that the fluctuations we observe initiate in the outer boundary layer, where the influence of the magnetic field is probably negligible.
Because the fluid is in rapid rotation beneath the outer sphere at rest, the outer boundary is very unstable, and subject to non-geostrophic instabilities.
We therefore expect a radically different behaviour when the outer sphere spins and the boundary layer is of Ekman type.

\subsection{Comparison with experimental results}\label{subsec:comparison_with_experiment}

\begin{figure*}
	\begin{center}
		\includegraphics[width=6cm]{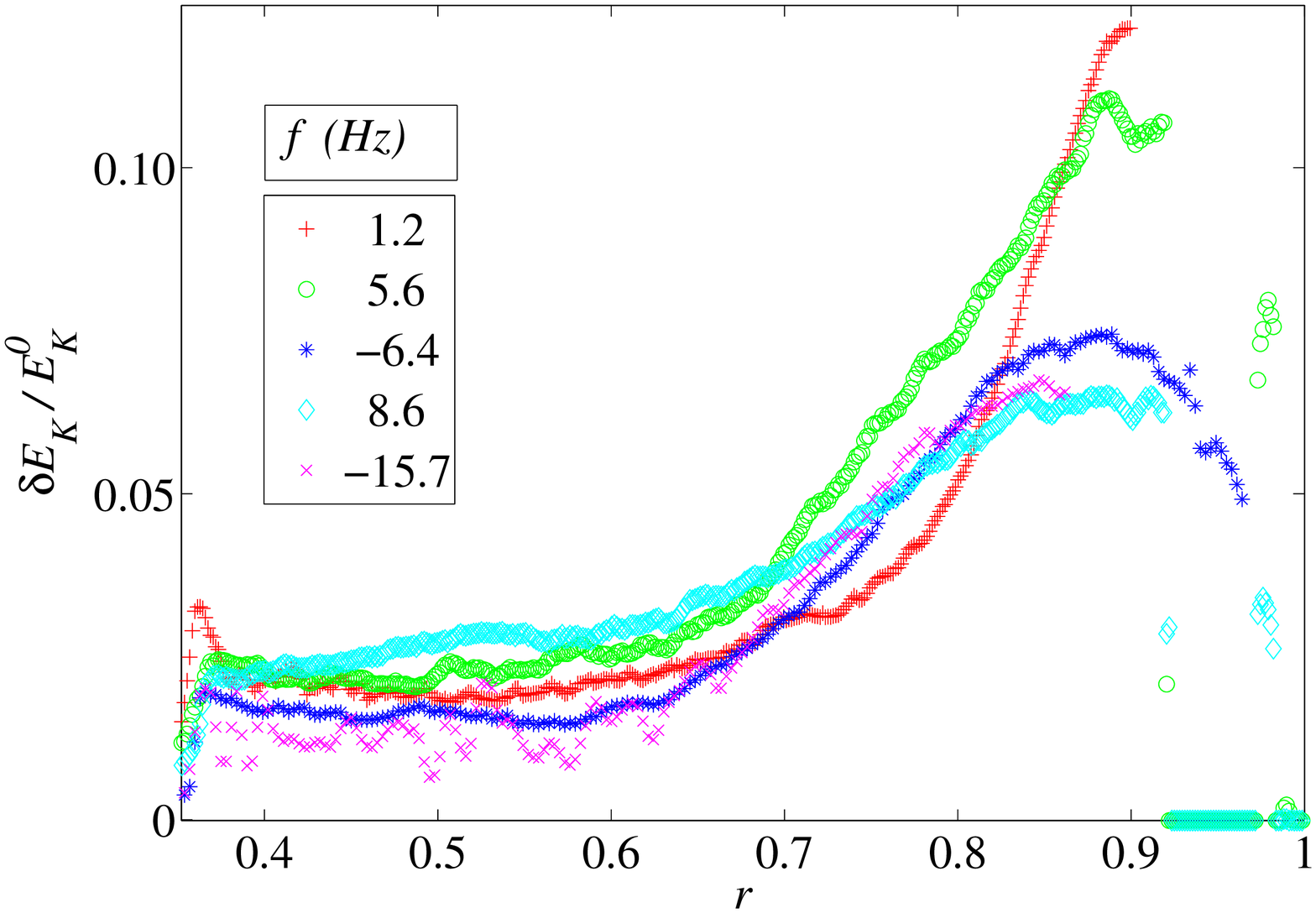}
		\includegraphics[width=6cm]{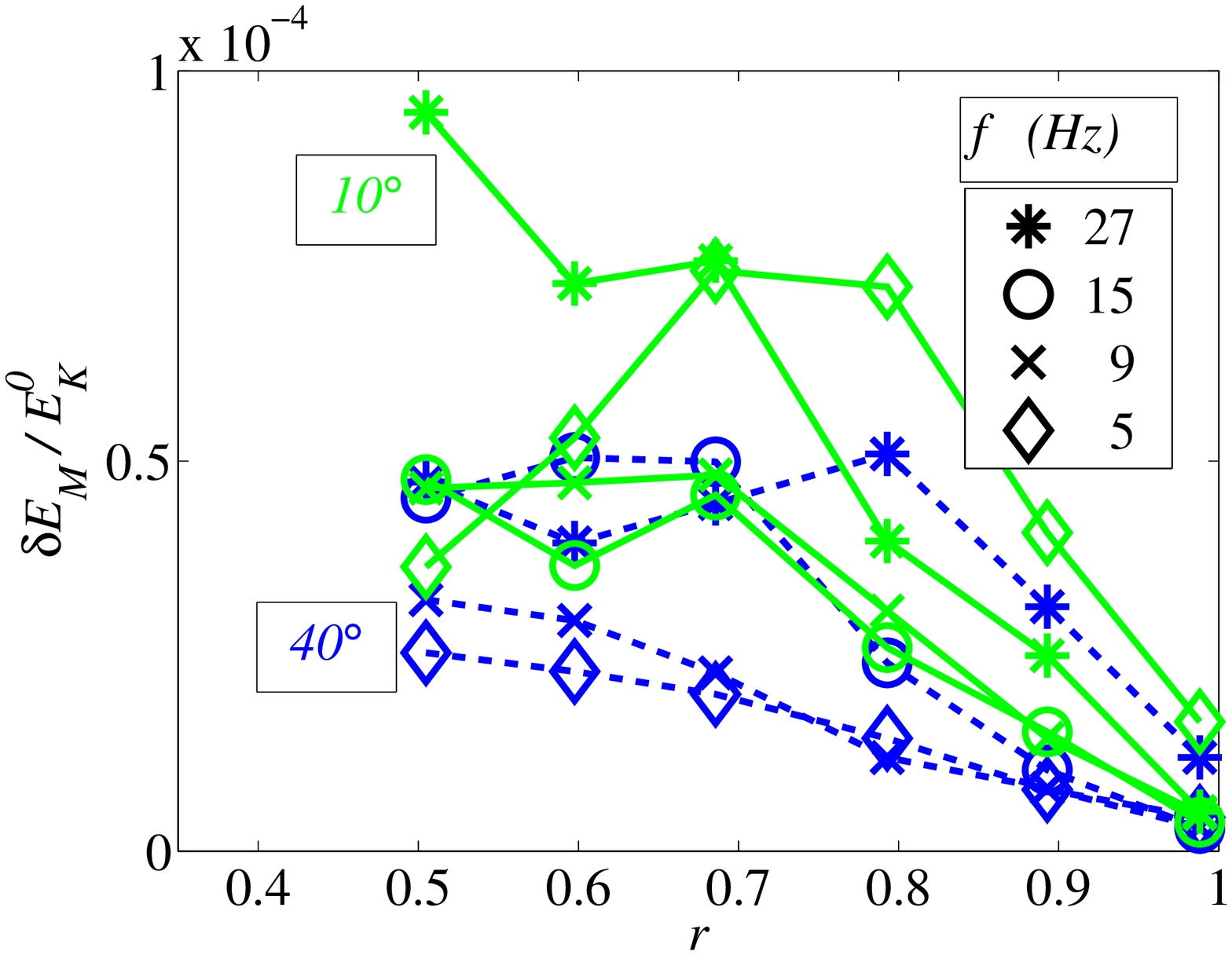}\\
		\includegraphics[width=10cm]{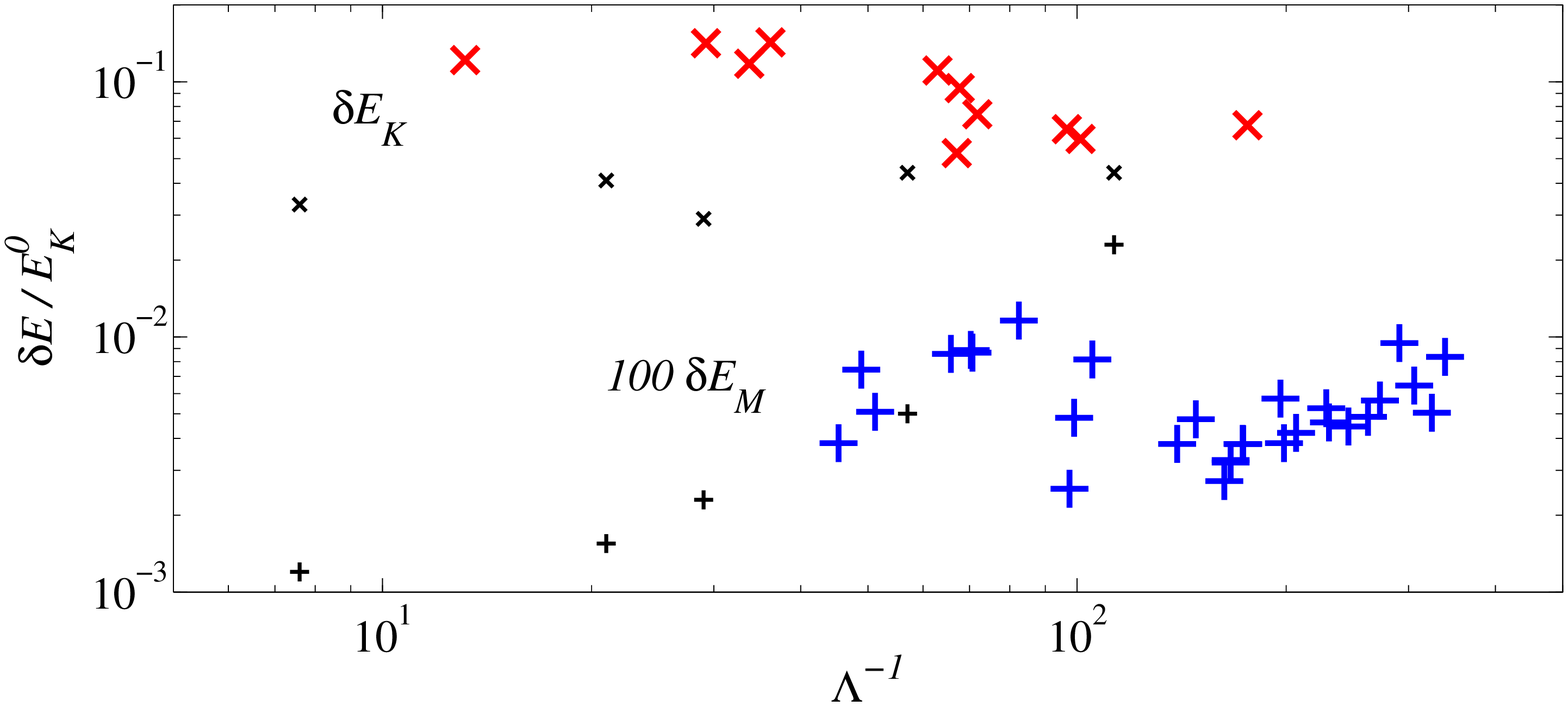}
	\end{center}
	\caption{Energy of the fluctuations in the $DTS$ experiment (all energy densities are normalized by $E_K^0 = \frac{1}{2} \rho \Omega^2 r_i^2$).
a) Selected radial profiles of the kinetic energy of the fluctuations $\delta E_K$ deduced from ultrasonic Doppler radial velocity measurements at latitude $-20^\circ$.
b) Selected radial profiles of the magnetic energy of the fluctuations $\delta E_M$ deduced from $b_\varphi$ measurements in a sleeve at two different latitudes ($10^\circ$ -solid green- and $40^\circ$ -dashed blue).
The symbols correspond to different rotation rates $f$ of the inner sphere, given in Hz in the legends.
c) Evolution of the kinetic (`x') and magnetic (`+') energies as a function of the inverse of the Elsasser number.
The black symbols are from the numerical simulations.
The colored symbols are deduced from the maximum value of the experimental radial profiles of $\delta E_K$ and $\delta E_M$ at various latitudes and forcings $f$, using the same markers.
Note that magnetic energies are multiplied by 100. 
}
	\label{fig:manips}
\end{figure*}

We cannot measure the total kinetic and magnetic energies of the fluctuations in the $DTS$ experiment.
However, we can get a quantitative assessment of the energy of the fluctuations as a function of radius, at given latitudes.
The kinetic energy is obtained from the fluctuations of the radial velocity measured by ultrasound Doppler velocimetry along a radial shot at a latitude of $-20^\circ$.
The magnetic energy is derived from the fluctuations of the azimuthal component of the magnetic field measured at three latitudes $10^\circ$, $20^\circ$ and $40^\circ$ and at 6 different radii, using Hall probes inserted in a sleeve, after removing a contribution at the rotation frequency $f$ and harmonics, which is due to small heterogeneities of the imposed magnetic field.
All energy densities are scaled with $E_K^0 = \frac{1}{2} \rho \Omega^2 r_i^2$.
As for the simulations, we integrate over azimuth by multiplying the measured $rms$ by $2 \pi r \sin\theta$ and convert to energy.
In order to relate to the numerical results, we assume that fluctuations are isotropic.
Additional measurements of the magnetic energy from radial and orthoradial probes partly support this hypothesis.
Nevertheless, the comparison remains approximative.

The kinetic energy profiles (figure \ref{fig:manips}a) confirm that fluctuations are strongest near the outer surface.
The maximum is deeper than in the numerical simulations (compare with figure \ref{fig:comp_delta_energy_u_b_Res_Has}a), a consequence of the much higher Reynolds number.
Note however that the thin viscous boundary layer cannot be resolved from the Doppler velocity profiles.

The magnetic energy profiles (figure \ref{fig:manips}b) clearly show that fluctuations are strongest near the inner sphere.
Figure \ref{fig:comp_delta_energy_u_b_Res_Has}b shows that only the simulation with the highest Reynolds number displays this behaviour.

Figure \ref{fig:manips}c compares the kinetic and magnetic energies of the fluctuations obtained from both the simulations (black crosses) and the experiments (colored symbols).
Since we don't have the full latitudinal dependence in the experiments, we simply take the maximum of each profile as an estimate of the overall energy.
The horizontal axis is  $\Lambda^{-1}$, the inverse of the Elsasser number.
It measures the ratio of the inertial force to the Lorentz force.
The magnetic energy remains much smaller than the kinetic energy.
It clearly increases with $\Lambda^{-1}$ in the simulations: velocity fluctuations penetrate deeper into the fluid and induce larger magnetic fluctuations because the imposed magnetic field is stronger there.
The experimental data follow the same trend for small forcing $f$, but there seems to be a strong drop near $\Lambda^{-1}=100$, before it increases again.
We have no explanation for this behaviour.

\subsection{The role of the Lorentz force}\label{subsec:Lorentz_force}

Although magnetic energies are much smaller than kinetic energies, the Lorentz force plays a major role.
The strong imposed dipolar magnetic field governs the dynamics of the mean flow in the $DTS$ experiment.
In particular, the very efficient entrainment of the fluid by the conductive inner sphere, and the zone of super-rotation next to it, are entirely due to the presence of the magnetic field.

\begin{figure*}
	\begin{center}
		\includegraphics[width=5cm]{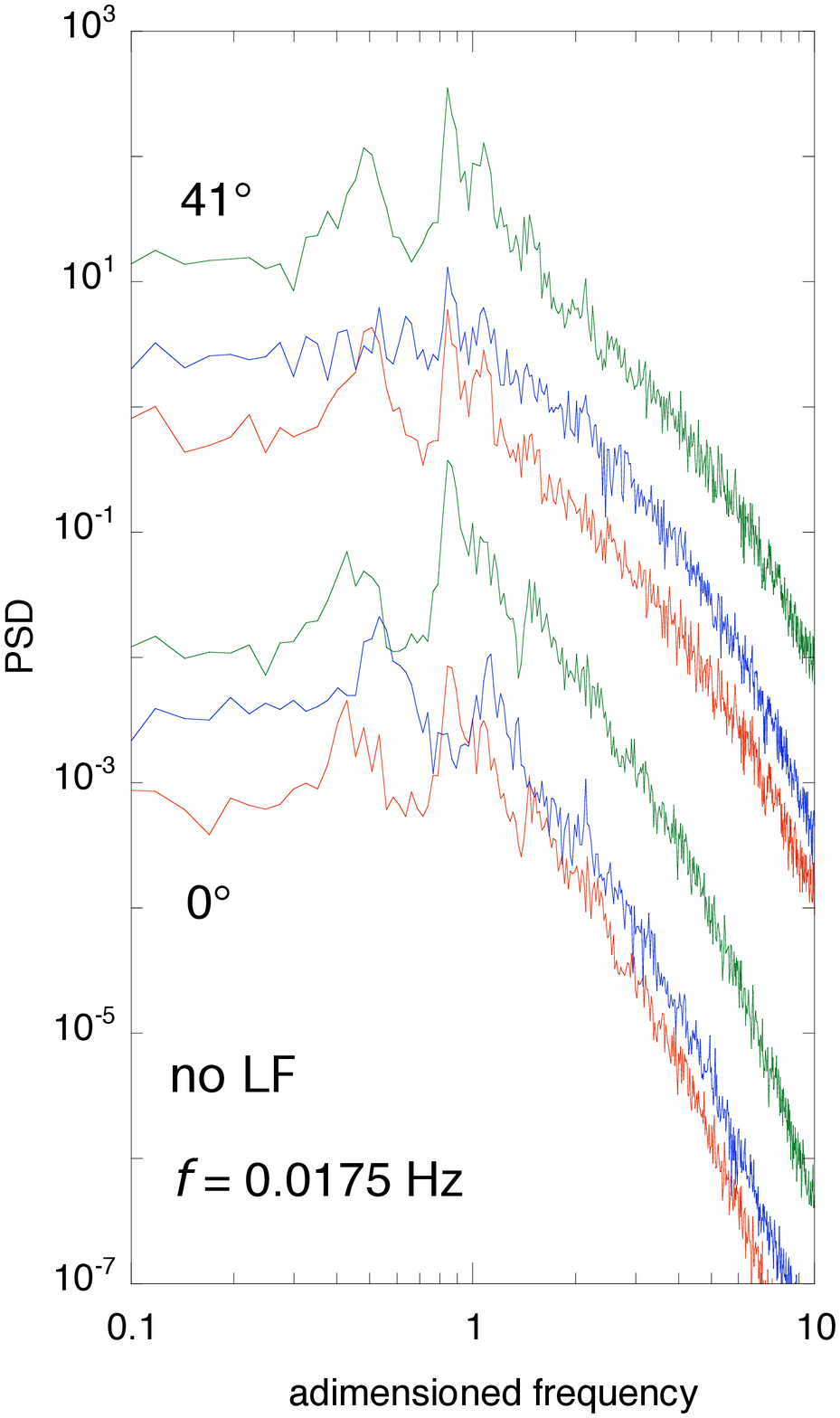}
		\includegraphics[width=6cm]{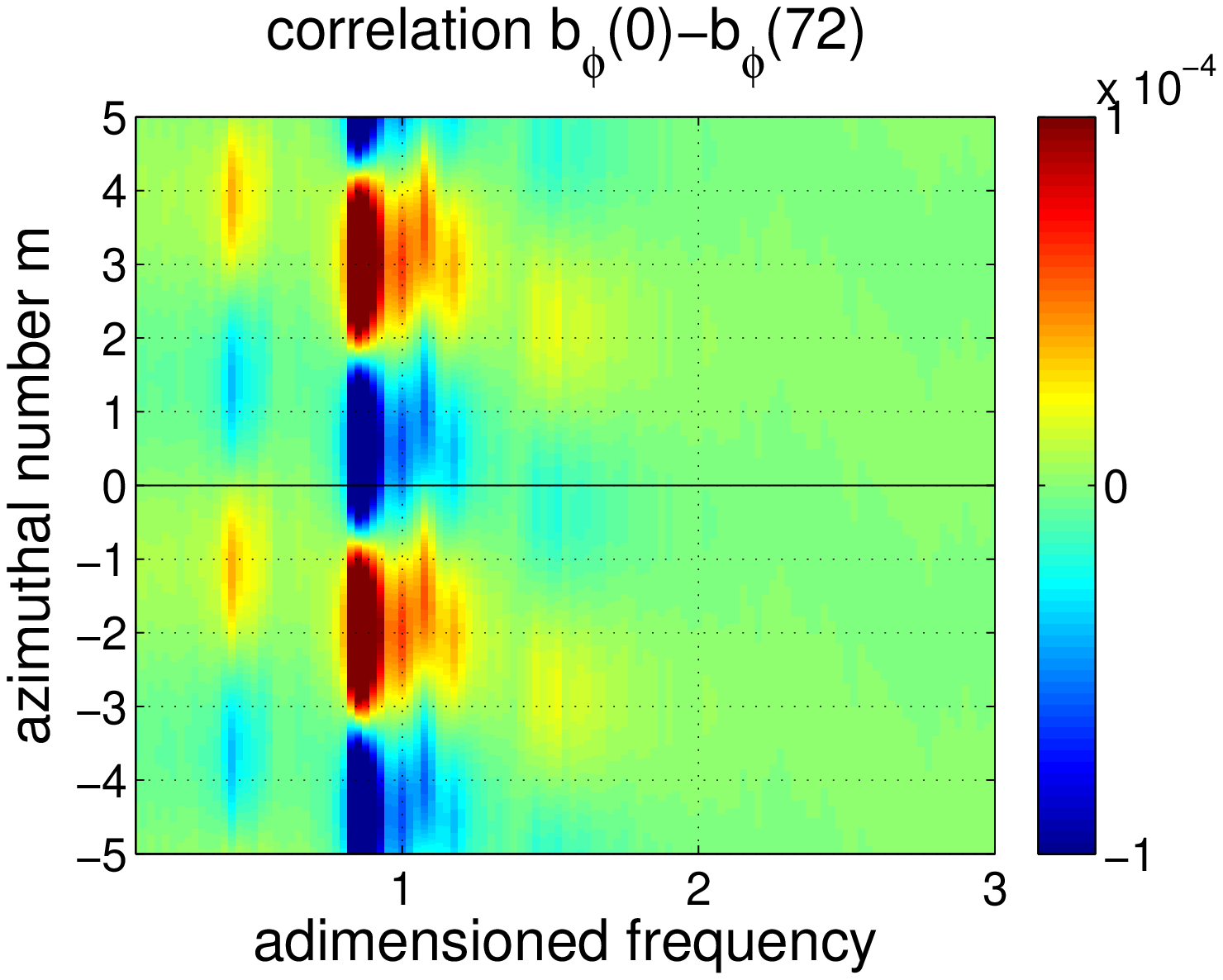}
	\end{center}
	\caption{Same as figure \ref{fig:spectra} but for a simulation where the Lorentz force is nulled out for $m \neq 0$.
Sharp spectral peaks are observed on all three components of the magnetic field. Neighbouring frequencies can have different azimuthal modenumber $m$.}
	\label{fig:spec_b_f_t_LM0_Re_2600_lat_long}
\end{figure*}

In order to see the effect of the Lorentz force on the fluctuations, we have run a simulation in which the Lorentz force is nulled out except for $m=0$.
We find that the radial profile of angular velocity at the equator remains essentially the same, illustrating that non-linear interactions of fluctuations with $m \neq 0$ barely contribute to the mean flow.
Frequency spectra of the surface magnetic field (figure \ref{fig:spec_b_f_t_LM0_Re_2600_lat_long}a) still display spectral bumps, but they are narrower and much more intense.
Furthermore, the azimuthal mode number analysis (figure \ref{fig:spec_b_f_t_LM0_Re_2600_lat_long}b) reveals that modes of a given $m$ show up at several distinct frequencies.

\begin{figure*}
	\begin{center}
		\includegraphics[width=6.5cm]{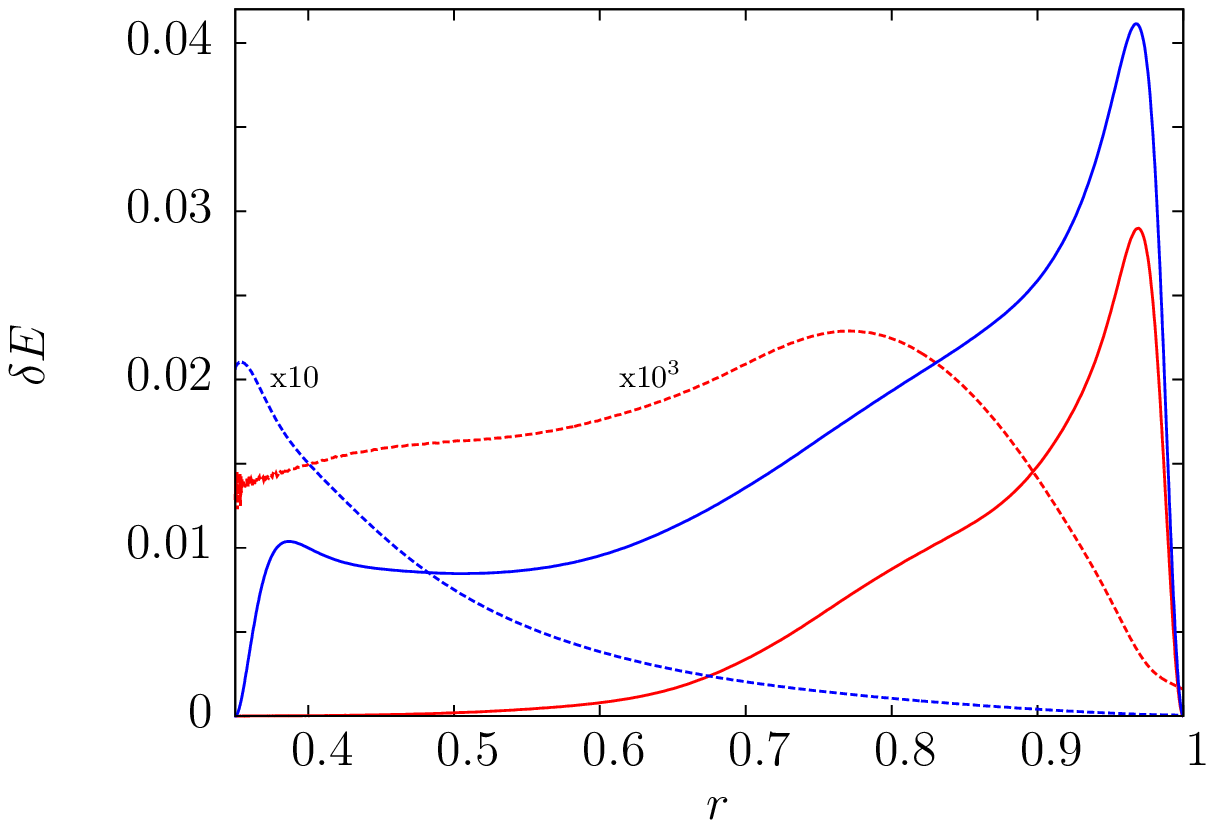}
	\end{center}
	\caption{Time-and-colatitude-averaged energy density of the fluctuations as a function of radius $r$.
This plot compares the kinetic (solid lines) and magnetic (dashed lines) energies of the reference simulation (in red) with a simulation (in blue) with the same parameters ($Pm=10^{-3}$, $\Rey=2\,611$ and $\Lambda=3.4 \times 10^{-2}$), but in which the Lorentz force has been nulled out for $m \neq 0$.
Note that the magnetic energy is two orders of magnitude larger in the latter case.
}
	\label{fig:delta_energy_u_f_r_Ms_M0_Re_2600}
\end{figure*}

The damping effect of the Lorentz force is best illustrated by plots of the radial profile of the energy of the fluctuations.
Figure \ref{fig:delta_energy_u_f_r_Ms_M0_Re_2600} displays the resulting radial profiles of the kinetic and magnetic energies, integrated over time and colatitude.
While the kinetic energy of the fluctuations is negligible for $r < 0.6$ when the Lorentz force is present, fluctuations invade the complete fluid shell when it is suppressed.
Fluctuations are largest beneath the outer shell where the magnetic field is weakest, but even there fluctuations are much weaker when the Lorentz force is active.

The magnetic energy of the fluctuations reaches only a thousandth of the kinetic energy in the $DTS$ configuration.
Not surprisingly, when the Lorentz force is suppressed, it jumps by a factor of 100 near the inner sphere, where the imposed magnetic field is strongest.
This demonstrates how dangerous it can be to infer magnetic energy and dissipation from flow solutions computed without the feedback of the Lorentz force (see discussion between  \citet{glatzmaier08} and \citet{Liu08}).

\section{Discussion}\label{sec:discussion}

We have obtained bumpy frequency spectra in numerical simulations of the magnetized spherical Couette flow (figure \ref{fig:spectra}a).
They compare very well with spectra obtained in the $DTS$ experiment from magnetic and electric time-records \citep{schmitt08}.
As in the experiment, the dominant azimuthal mode number is incremented by 1 from one frequency bump to the next (figure \ref{fig:spectra}b).
Very large Reynolds numbers are not required for getting this behaviour, but one needs to accumulate long time-series (typically 300 turns of the inner sphere) for the peaks to show up clearly in the spectra.
However, we note that bumps in the numerical spectra are not as pronounced as in the experiments.

We have developed a new method, which bridges the gap between the linear modal approach \citep{Rieutord97, Kelley07, schmitt08} and full non-linear simulations and experiments.
By performing a time-domain Fourier transform of the full fields for each azimuthal number $m$, we recover the dominant frequencies (figures \ref{fig:fft_spectra_1} and \ref{fig:fft_spectra_2}) and obtain the structure of the modes (figure \ref{fig:fft_modes}), which can then be compared to linear solutions and to experimental observations.
We think that this approach will help identifying the mode selection mechanism in other experiments \citep{kelley10}.

Snap-shots (figure \ref{fig:added}) and spatio-temporal plots (figure \ref{fig:spatio_temporal}) reveal a rather different story, in which chaotic instabilities are swept by the flow.
We could show that these two views are dual: since the instabilities circle around the sphere, the parts that are in phase between two successive passages are statistically enhanced.
Since the spectral bumps are more pronounced in the experiments, this effect appears to be more efficient at large Reynolds number.

The maps of the kinetic energy of the fluctuations (figures \ref{fig:delta_energy_u_f_r_theta_Re_2600}a, \ref{fig:comp_delta_energy_u_b_Res_Has}a and \ref{fig:manips}a)  indicate that they initiate in the outer boundary layer, with only minor influence of the magnetic field.
Instabilities appear above a critical Reynolds number $\Rey_c=1 \, 860$.
We identify two types of instabilities: i) axisymmetric polewards migrating rolls at high latitude; ii) non-axisymmetric ($m=2$ at the threshold) secondary instabilities of an equatorial centripetal jet.
The first type is similar to the instabilities of a B\"odewadt layer.
The second type resembles the jet-instability of the centrifugal equatorial sheet in non-magnetized spherical Couette flow.
The two instabilities are coupled by the meridional circulation (they trigger one another), and the system quickly evolves towards a chaotic state in which outer boundary layer instabilities are swept around by the azimuthal and meridional large scale flows.


\citet{schmitt08} observe that fluctuations in $DTS$ are delayed and reduced when the outer sphere is spinning.
We think that this is because the boundary layer is then closer to an Ekman-type, which is much more stable \citep{Lingwood97}, and that non-geostrophic instabilities are hampered.

The fluctuations of kinetic energy are much larger than that of magnetic energy (figure \ref{fig:manips}c), which are mostly slave of the former.
If we assume that the magnetic fluctuations scale as $b \simeq a Rm B$, we obtain that the ratio of the magnetic to kinetic energy behaves as $\delta E_M / \delta E_K \simeq a^2 Lu^2$, where $Lu$ is the Lundquist number defined in section \ref{subsec:dimensionless_parameters}.
Since the Lundquist number is of order 1 in both the experiments and the simulations the $a$ pre-factor must be rather small to explain the observed energy contrast.
As a matter of fact, direct measurements of the mean induced azimuthal magnetic field yield $a \simeq 0.1$.
Both the $DTS$ measurements (figure \ref{fig:manips}b) and our largest Reynolds number numerical simulation (figure \ref{fig:comp_delta_energy_u_b_Res_Has}b) display a strong increase of the magnetic energy fluctuations when getting closer to the inner sphere.
This appears to be essentially the consequence of the strong increase of the imposed dipolar field there.
Indeed, the local Lundquist number increases from $Lu=0.5$ at the equator of the outer sphere to $Lu=12$ at the equator of the inner sphere.

At first order, we expect both energies to be proportional to the square of the imposed inner sphere velocity.
However, we note that when scaled accordingly, the kinetic energy tends to decrease when the forcing is increased, while the scaled magnetic energy increases (figure \ref{fig:manips}c).
\citet{Brito11} showed that the energy of the mean flow behaves similarly, and proposed that this is a consequence of the increasing turbulent friction at the outer surface: as the friction increases, the core-flow is slowed down, while the shear between the spinning inner sphere and the fluid increases, inducing a stronger magnetic field. 
We think that another effect explains the trend observed for the energy fluctuations: as the forcing increases, the damping effect of the magnetic field decreases.
Instabilities penetrate deeper into the fluid and produce larger magnetic fluctuations, even though their scaled kinetic energy is reduced because of the decreased velocity drop across the outer boundary layer.
We note that for large forcing $f$ the magnetic energy is smaller in the experiments than in the simulations, suggesting again the role of the strong turbulence.

Although the magnetic energy is very small, the Lorentz force plays a major role: it determines the very efficient entrainment of the fluid by the spinning inner sphere, but it also heavily damps the fluctuations in most of the fluid.
When we remove the Lorentz force for $m \neq 0$, fluctuations invade the fluid (figure \ref{fig:delta_energy_u_f_r_Ms_M0_Re_2600}), and sharper and more numerous frequency peaks are observed in the spectra (figure \ref{fig:spec_b_f_t_LM0_Re_2600_lat_long}).
This gets closer to the observations of \citet{Kelley07}, where the imposed magnetic field was weak and only served as a marker of the flow.

Even though bumpy frequency spectra are observed in both situations (weak or strong magnetic field), they differ in several important aspects.
In the $DTS$ experiment, we observe broad peaks corresponding to azimuthal mode numbers up to $m=10$ for all rotation rates $f$ of the inner sphere, when the outer sphere is at rest.
Both equatorially symmetric and anti-symmetric modes are present \citep{schmitt12}.
The fluid is efficiently entrained by the magnetic coupling with the spinning inner sphere, and the largest velocity gradients are located near the outer boundary.
Modes and fluctuations are strongly damped by the imposed magnetic field. 

In contrast, when the magnetic field is weak and the outer sphere is also spinning, as in \citet{Kelley07} and \citet{rieutord12}, the spectra are dominated by sharp peaks corresponding to inertial modes with selected azimuthal numbers $m$.
Only equatorially anti-symmetric modes appear to be excited \citep{rieutord12}.
Most of the fluid rotates rigidly with the outer sphere, and velocity gradients are strong only in the Stewartson layer tangent to the inner sphere.
\citet{rieutord12} show that this layer can behave as a critical layer, thereby exciting modes with a dominant azimuthal mode number $m=-4 \hat{\omega} / Ro$, where $\hat{\omega}$ is the frequency of the mode.
The Rossby number is defined as $Ro=f_i/f_o-1$, where $f_i$ and $f_o$ are the rotation frequencies of the inner and outer spheres, respectively.

Because of these differences, we don't expect the mechanism proposed by \citet{rieutord12} to apply to the situation discussed in this article.
However, it would be interesting to investigate whether the idea of critical layers can help understanding the sort of statistical resonance we invoke to explain our observations.

Note that in our geometry, one could have expected instabilities to develop in the inner region near the equator, where the flow obeys Ferraro law, and where a small velocity perturbation produces a large Lorentz force.
This does not appear to be the case.
In the present study, we have kept the Lundquist number small, as in the $DTS$ experiment.
Alfv\'en waves are therefore damped out rapidly.
They might still contribute to shaping the modes near the inner sphere.
It would be interesting to investigate the turbulent regime in the $DTS$ geometry at larger Lundquist number.

\vspace{6pt}
We thank D. Jault for stimulating discussions.
R. Hollerbach and three other referees helped us improve our manuscript.
We gratefully acknowledge the support of CNRS and Universit\'e de Grenoble through the collaborative program ``Turbulence, Magnetohydrodynamics and Dynamo''.
Part of the numerical simulations were run at the Service Commun de Calcul Intensif de l'Observatoire de Grenoble (SCCI).

\vspace{6pt}
Supplementary movies are available at journals.cambridge.org/flm.

\vspace{6pt}
Movie 1. Map view of the time evolution of the azimuthal velocity $u_\varphi$ beneath the surface of the outer sphere ($r = 0.95$) in the reference numerical simulation ($Pm=10^{-3}$, $\Rey=2\,611$ and $\Lambda=3.4 \times 10^{-2}$).
At the origin time, the fluid is at rest and the rotation rate of the inner sphere is set to $f$.
Time $t$ is measured in rotation periods of the inner sphere.
There are 6 frames per turn and the movie lasts 100 turns.
Note that the first instabilities appear at the equator and are axisymmetric.
After about 24 turns, non-axisymmetric instabilities show up.

\vspace{6pt}
Movie 2. Time evolution of the angular velocity $\omega =  u_\varphi /s$ in a meridional plane ($\varphi = 0$) for the same simulation as in movie 1.

\bibliographystyle{jfm}

\bibliography{biblio}

\end{document}